\documentclass[1,12pt]{elsarticle}

\usepackage{graphicx}
\usepackage{color}
\usepackage{xcolor}
\usepackage{textcomp}
\usepackage{gensymb}
\usepackage{amsmath}
\usepackage{tabularx}
\usepackage{subcaption}
\usepackage{array}
\usepackage{amssymb}
\usepackage{lineno}
\usepackage{float}
\usepackage{booktabs}
\usepackage{soul}

\biboptions{comma,square,sort&compress}

\journal{Acta Materialia, accepted for publication.}

\begin{document}
\sloppy 
\begin{frontmatter}

\title{An analysis of the influence of the precipitate type on the mechanical behavior of Al - Cu alloys by means of micropillar compression tests}

 \author{B. Bell\'on$^{1, 2}$}
\author{S. Haouala$^{1}$}
\author{J. LLorca$^{1, 2, }$\corref{cor1}}
\address{$^1$ IMDEA Materials Institute, C/ Eric Kandel 2, 28906, Getafe, Madrid, Spain. \\  $^2$ Department of Materials Science, Polytechnic University of Madrid/Universidad Polit\'ecnica de Madrid, E. T. S. de Ingenieros de Caminos. 28040 - Madrid, Spain.}

\cortext[cor1]{Corresponding author}

\begin{abstract}

The influence of different types of precipitates (either Guinier-Preston zones, $\theta''$ or $\theta'$) on the critical resolved shear stress and strain hardening was determined by means of micropillar compression tests in an Al - 4 wt. \% Cu alloy. The size, shape and volume fraction of the precipitates were measured in each case. It was found that size effects were negligible for micropillars with diameter $\ge$ 5 $\mu$m. Micropillars with Guinier-Preston zones showed strain localization due to precipitate shearing. The best mechanical properties were obtained with either a fine dispersion of the $\theta''$ precipitates or a coarser dispersion of $\theta'$. Both precipitate shearing and Orowan loops were observed around the $\theta''$ precipitates and the micropillar strength was compatible with the predictions of the Orowan model. In the case of the alloy with $\theta'$ precipitates, the strengthening contribution associated with the transformation strain around the precipitates has to be included in the model to explain the experimental results. Finally, the micropillar compression tests in crystals with different orientations were used to calibrate a phenomenological crystal plasticity. This information was used to predict the mechanical properties of polycrystals by means of computational homogeneization. 
\end{abstract}


\end{frontmatter}

\section{Introduction}
\label{S:1}

Cu is one major alloying elements in most wrought Al alloys that are widely used in transportation (aerospace, automotive, railways, marine, etc.) because of the low density, limited cost, ease of fabrication and excellent combination of mechanical properties (strength, ductility and toughness) \cite{Polmear2017}. Addition of Cu atoms to Al increases the strength due to solution and precipitation hardening \cite{Polmear2017,Nie2014,Argon2008}. The contribution of solution hardening is limited by the solubility of Cu in Al while precipitate hardening can be tailored by controlling the geometric features of the precipitates (size, shape, spatial distribution and volume fraction) through thermo-mechanical treatments. Moreover, the precipitate size as well as the nature of the matrix/precipitate interface (coherent, semi-coherent or incoherent) determine the mechanisms of dislocation/precipitate interactions and, thus, the strengthening contribution \cite{N97}.

Because of its industrial relevance and the high strength achieved, precipitation hardening in the Al-Cu system has been widely studied. The types of metastable precipitates that appear during thermal treatments of a supersaturated solid solution of Cu atoms have been established from careful experimental analysis \cite{Argon2008, Nie2014, Ringer2000,Son2005}. It begins with the formation of Guinier-Preston (GP) zones at ambient temperature \citep{Guinier1938,Preston1938}, which stand for circular disks made of one layer of Cu atoms coherent with the Al matrix, that grow parallel to the \{100\} faces of the Al FCC lattice. If the alloy is aged at high temperature (from 130 to 190 $^o$C) \cite{Silcock1953,Nie2014}, $\theta''$ (Al$_3$Cu) precipitates, which also grow as disks with the same orientation, appear instead of GP zones. They have a  FCC structure obtained by replacing every fourth layer of the Al atoms by Cu atoms on (001) planes and are coherent with the matrix. $\theta''$ are eventually replaced by $\theta'$ (Al$_2$Cu) precipitates with a BCT structure and disk shape. The broad face of the $\theta'$ disks --parallel to the \{100\} faces of the Al FCC lattice-- is coherent with the matrix while the edge are semi-coherent. Finally, the $\theta$ (Al$_2$Cu) precipitate has a complex BCT structure, incoherent with the matrix and different shapes and orientations with respect to the matrix have been reported \cite{Kaira2018, VS67}. 

The precipitation sequence and kinetics in Al-Cu alloys have been amply studied by transmission electron microscopy (TEM) \cite{Boyd1971,Ringer2000,Son2005,Bourgeois2011, Alex2018,Biswas2011} and the experimental data have been corroborated by recent multiscale simulations based on first principles calculations in combination with either Kinetic Monte Carlo or phase-field modelling \cite{KPG17, Liu2017, LPL19, SH19, MKI19}. Only GP zones are formed at ambient temperature while precipitation at high temperature begins with the homogeneous nucleation of the $\theta''$ phase followed by the heterogeneous nucleation of $\theta'$ upon dislocations afterwards. The $\theta''$ precipitates grow and eventually disappear with longer aging time, while homogenous precipitation of  $\theta'$ is observed. Finally, the $\theta$ phase appears (initially on grain boundaries or at $\theta'$/Al interfaces after longer aging times if the temperature is sufficiently high (above $>$ 200$^\circ$C) \cite{WO01, LPL19}.

Due to complex precipitate sequence, strengthening of Al-Cu often involves the contribution of different metastable precipitates which present different interactions with the dislocations. Most authors agree that GP zones are sheared by dislocations \cite{Byrne1961, Nie2014} in agreement with recent atomistic simulations \cite{EMS19, EBM20} while  $\theta'$ precipitates are by-passed by the formation of Orowan loops \cite{DaCostaTeixeira2008, Nie2008,Sehitoglu2005, Ma2016, Li2017, Kaira2019}. However, the dislocation/precipitate interaction mechanisms in the case of $\theta''$ precipitates are not known. Moreover, the strengthening provided for each type of precipitate is not known either because most of the studies on the mechanical properties of Al-Cu alloys have been carried out in polycrystalline samples (see, among many others, \cite{Calabrese1974, Calabrese1974b, Merle1981a, DaCostaTeixeira2008, DaCostaTeixeira2009, Chen2013}) and it is very difficult to isolate the precipitate contribution from those coming from grain boundaries and solid solution. A limited number of investigations have tried to overcome these limitations by testing Al-Cu single crystals \cite{Byrne1961, Price1964, Russell1970, Muraishi2002, Sehitoglu2005} but the emphasis was placed in understanding the strain hardening mechanisms in different orientations as a function of the type of precipitate. Moreover, the mechanical characterization was not accompanied by a quantitative analysis of the type and geometrical characteristics of the precipitates and, thus, no information could be obtained about what precipitate provides higher strengthening. 

In this investigation, the dislocation/precipitate interactions of $\theta''$ precipitates and the strengthening provided by different types of precipitates (either GP zones, $\theta''$ or $\theta'$) was measured along different orientations by means of micropillar compression tests in single crystals. The size, shape and volume fraction of each type of precipitates was carefully determined by means of TEM and this information, in combination with the results of the mechanical tests, can be used to validate the models of precipitation hardening available in the literature \cite{Santos-Guemes2018, EMS19, EBM20}. In addition, the results of the mechanical tests in orientations suitable for single, double and multiple slip were used to calibrate a phenomenological crystal plasticity model that can be used to predict the behavior of polycrystal containing different types of precipitates.

The paper is organized as follows. The experimental details of the processing, microstructural and mechanical characterization are presented in section 2. The experimental results for the alloys containing different types of precipitates are detailed in section 3, while the crystal plasticity model and the calibration with the mechanical tests in micropillars is presented in section 4, together with the predictions for the polycrystal behavior. Finally, the conclusions of the paper are summarized in section 5.

\section{Material and experimental procedures}
\label{S:2}

\subsection{Processing and microstructure}\label{SS:Processing}

An Al-4 wt.\% Cu alloy was prepared using high-purity metals by casting in an induction furnace (VSG 002 DS, PVA TePla). Samples were subjected to a homogenization and solution heat treatment during 22 h at $540^{\circ}$C, followed by quenching in water, leading to specimens with an average grain size above several hundreds $\mu$m. Afterwards, some samples were aged at ambient temperature (natural aging) and others were subjected to artificial aging at 180$^{\circ}$C during 30 and 168 hours. 

The precipitate structure was carefully characterized by means of TEM (TEM FEI Talos). To this end, foils were ground to a thickness of about 100 $\mu$m and punched to produce circular discs of 3 mm in diameter. These foils were jet electropolished by using a chemical solution of 30 vol.\% nitric acid in methanol at $\approx$ -30$^{\circ}$C using a voltage of 15 V. The structure, dimensions, spatial distribution and volume fraction of the precipitates were determined in the naturally aged (NA) and artificially aged (AA30 and AA168) samples. Automatic processing of the TEM was difficult because of the diffuse contrast  between matrix and precipitates in some cases (particularly for small GP zones). Thus, the precipitates in one orientation variant of each micrograph were manually colored and  a simple binary color threshold was used to binarize the images. The number of precipitates, the apparent diameter and thickness were measured using the particle analyzer in ImageJ \cite{ImageJ12}. Then, the process was repeated for the precipitates in the other orientation variant in the image. The actual diameter and thickness distributions  were calculated following the procedure described in \citep{Nie2008, Alex2018}, taking into account that only 2 precipitate variants are visible in the TEM micrographs oriented along the $[001]_{\alpha}$ zone axis. The thickness of the foils in the beam direction was determined by measuring the spacing of Kossel- M\"ollnstedt fringes in the \{022\}$_\alpha$ in a $<100>_\alpha$ two-beam convergent beam electron diffraction pattern \cite{KJB75}.

\subsection{Micropillar compression}

\begin{figure}[h!]
     \centering
         \includegraphics[width=0.7\textwidth]{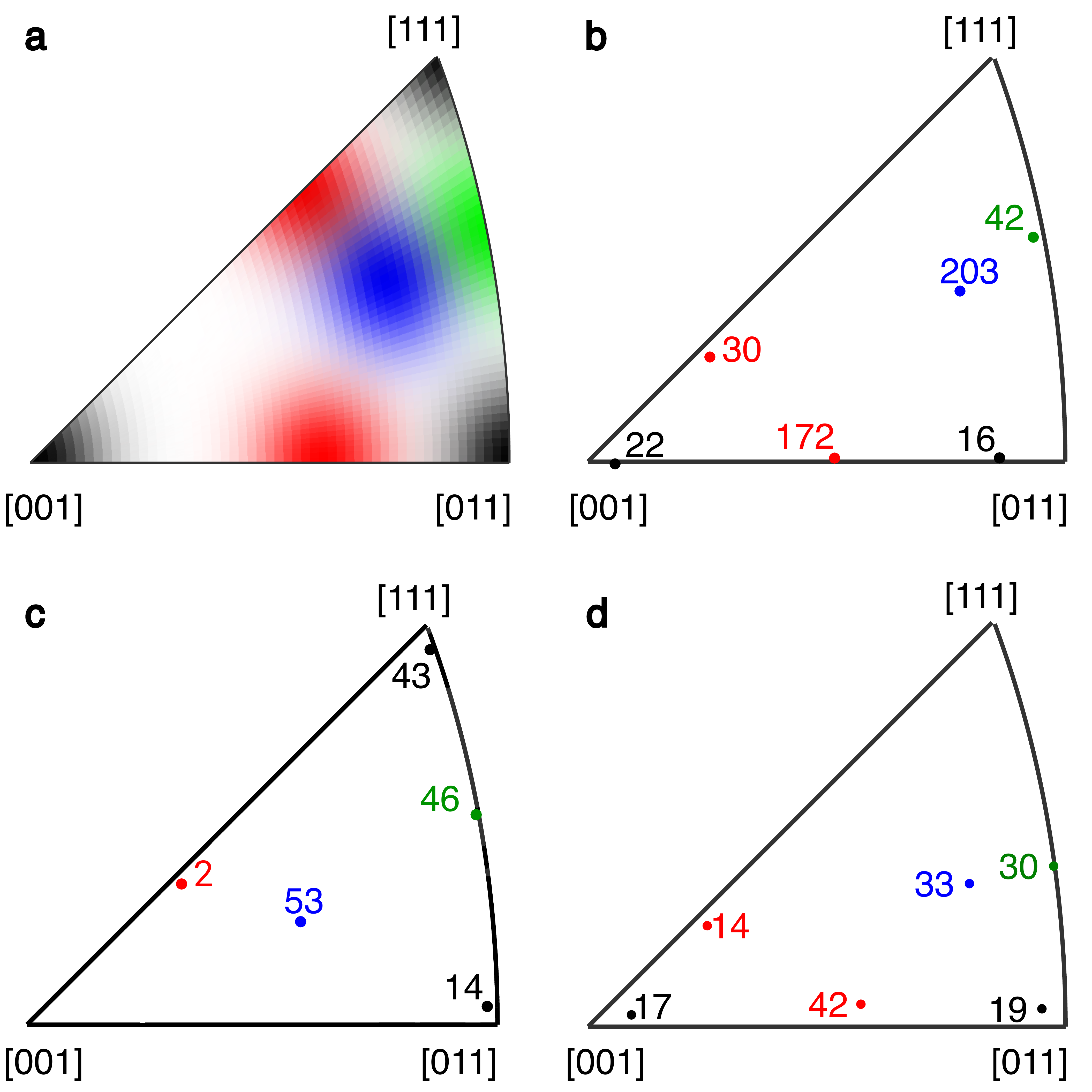}
 \caption{Inverse pole figures showing the orientation of the grains selected to carve micropillars in the samples subjected to different heat treatments. (a) Regions of the inverse pole figure for single slip (blue), double coplanar slip (green), double non-coplanar slip (red) and multiple slip (black). (b) Naturally aged. (c) Artificially aged for 30 hours at 180$^{\circ}$C.  (d) Artificially aged for 168 hours at 180$^{\circ}$C. The grain number is indicated along the dot indicating the orientation.}
        \label{fig:Orientations}
\end{figure}

The orientation of the grains to mill the micropillars with the desired orientation was determined by means of electron backscatter diffraction (EBSD) in a field-emission scanning electron microscope (FEI Helios NanoLab 600i). Orientation maps were analyzed using the MTEX toolbox \cite{MTEX} to filter, denoise and obtain the  Schmidt factors for each slip system in each grain, assuming that the applied stress was perpendicular to the sample surface. The orientation of the grains selected for micropillar compression is displayed in the inverse pole figures depicted in Fig. \ref{fig:Orientations}. It includes grains oriented for single slip (blue region), double coplanar slip (green), double non-coplanar slip (red) and multiple slip (black). Micropillars were carved in two different grains for each ageing condition and orientation. Between 4 to 5 micropillars were carved from each grain. Nevertheless, only the mechanical properties obtained in the micropillars carved from one grain will be shown below for the sake of brevity. It should be noted that the experimental scatter in the mechanical tests was limited and that no measurements were discarded except those which presented very low values of the initial stiffness of the micropillar, indicating a strong misalignment between the flat punch and the micropillar surface.

It has been reported in literature a strong increase in the flow stress of FCC single crystal metals as the micropillar diameter decreased \cite{Greer2011,Soler2012, Wharry2019}. The effect of micropillar diameter was ascertained for each precipitate distribution by means of mechanical tests in micropillars with circular cross sections and different diameters (in the range 2 to 7 $\mu$ m). They were machined from the grains oriented for single slip by focused ion beam (FIB) milling using Ga$^+$ ions at 30 kV in a dual-beam FEI Helios NanoLab 600i microscope. The initial current was 47 nA to remove the material around the pillar and provide sufficient clearance for the flat punch indenter. Milling continued with 9 nA followed by 2.5 nA to reduce the size of the pillar to the final dimensions. Currents of 0.79 and 0.43 nA were used in the steps to reduce the tapper of the pillar and minimize damage on the pillar surface.

Micropillars with the minimum dimensions to obtain size-independent values of critical resolved shear stress were then used to determine the strengthening provided by the different precipitate structures (either Guinier-Preston zones, $\theta''$ or $\theta'$ precipitates) in samples oriented for single, double and multiple slip (Fig. \ref{fig:Orientations}). The pillars had a square cross section of  5 $\mu $m x  5 $\mu $m and aspect ratios in the range 2.0-3.0. They were carved using the FIB with Ga$^+ $ ions operated at 30 kV, with currents starting on 47 nA to  provide sufficient clearance for the flat punch indenter. Milling continued with 9 nA and 2.5 nA to reduce the size of the pillar. Finally tapper was reduced at currents of 0.79 and 0.43 nA. 

Circular micropillars with different diameter were deformed using a high-load transducer (up to 900 mN) that can deform the micropillar with the largest diameter. Square micropillars with a cross-section of 5 $\times$ 5 $\mu$m$^2$ were  deformed with a low-load transducer (12 mN). The high load transducer was able to carry out the tests under real displacement control while the load-low transducer applied the displacement using a PID control. The taper angle was small ($<$ 2$^\circ$) for circular micropillars and negligible ($<$ 1$^\circ$) for the square micropillars. In all cases, the micropillars were milled from the center of the grains to ensure that they were single crystals.

Compression of the micropillars was carried out using a circular diamond flat punch of 10 $\mu$m of diameter using a nanoindenter (Hysitron TI950). Tests were carried out in displacement control at an average strain rate of $ \approx 10^{-3}$ s$^{-1}$ up to 10\% strain. The engineering stress-strain curves were calculated from the applied load using the upper cross-sectional area and length of the pillars, measured after the milling process. The engineering stress was transformed into the resolved shear stress (RSS) using the Schmid factor of the most suitable oriented slip system in the case of micropillars oriented for single slip. The critical resolved shear stress (CRSS) - which correspond to the beginning of plastic deformation - was determined  from the RSS-strain curves at a plastic strain of 0.02$\%$ \cite{Proudhon2008, ASMHandbookV2}.

The deformed micropillars were observed in the scanning electron microscope to analyze the deformation mechanisms from the shape of the micropillar and the surface traces. Moreover, thin foils parallel to the (001)$_\alpha$ and (100)$_\alpha$ planes were extracted after deformation from one sample aged at 180$^\circ$ during 30 hours using FIB and the dislocation/precipitate interactions were analyzed in the TEM.

\section{Experimental Results}\label{S:4}

\subsection{Precipitate distribution}

\begin{figure}[h!]
	\centering
	\includegraphics[width=\textwidth]{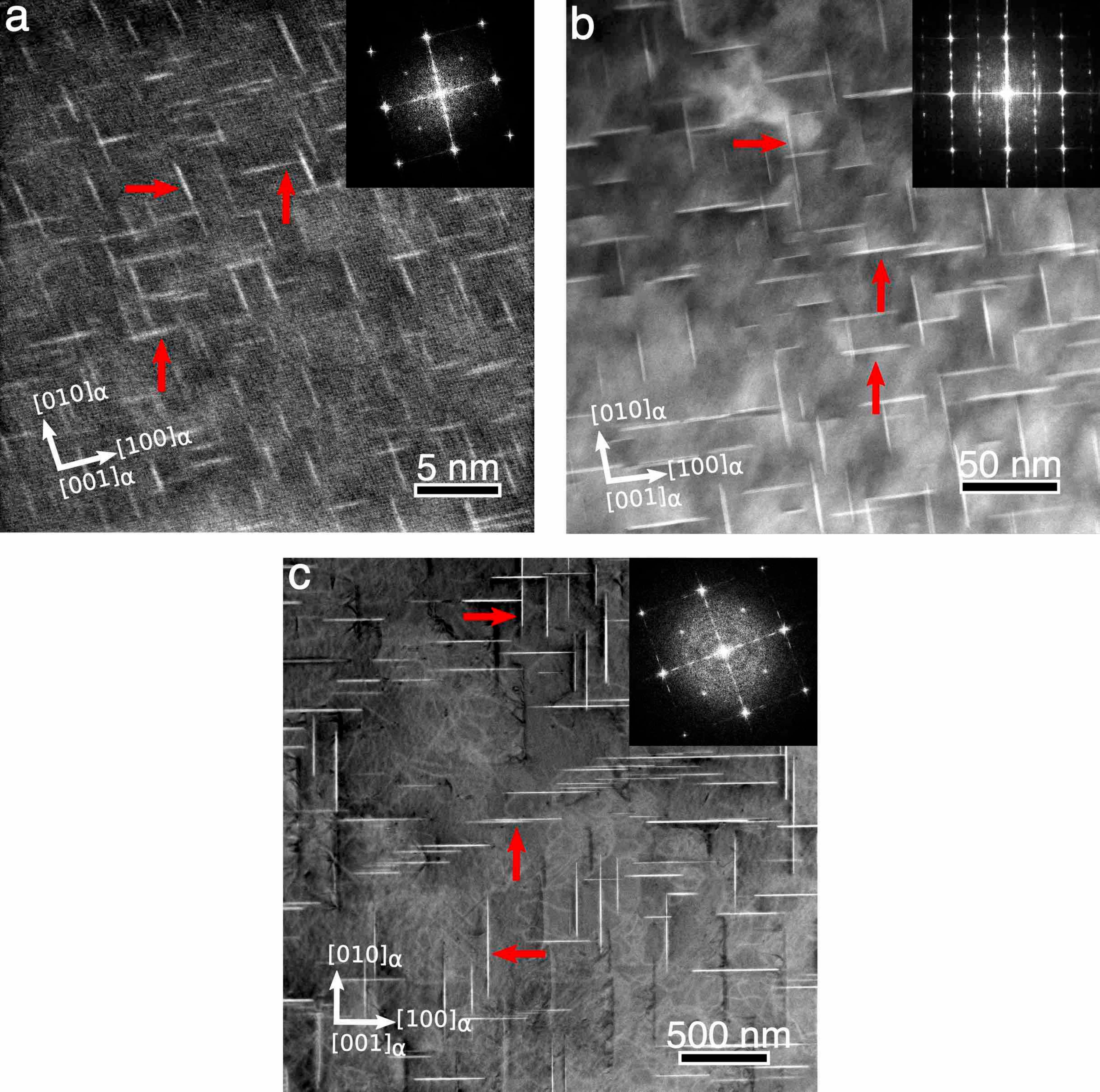}
	\caption{High-angle annular dark-field scanning-transmission electron micrographs of the Al-4 wt.\% Cu specimens, showing the precipitate distribution in the samples subjected to different heat treatments.  (a) GP zones in the sample naturally aged at ambient temperature. (b) $\theta''$ precipitates in the sample aged at 180$^\circ$C during 30 hours. (c) $\theta'$ precipitates in the sample aged at 180$^\circ$C during 168 hours. The zone axis was $[001]_{\alpha}$. The FFT of each type of precipitates are shown in the insets.}
	\label{fig:precipitados}
\end{figure}

Data about the dimensions of the precipitates were obtained from measurements in over about 300 precipitates in the TEM micrographs for each aging condition. The NA sample contained a homogeneous distribution of GP zones  (Fig. \ref{fig:precipitados}a). The main precipitates found in AA30 samples were $\theta''$, which were also homogeneously distributed in the microstructure (Fig. \ref{fig:precipitados}b). Clusters of $\theta'$ precipitates forming staircase structures were also found in this sample \cite{Alex2018}. These clusters developed because the  nucleation of $\theta'$ precipitates on dislocations is favoured by the interaction energy between the stress-free transformation strain around the precipitate and the dislocation stress field \cite{Liu2017}. However, these clusters of $\theta'$ precipitates did not contribute to the strengthening of the micropillars due to the small initial dislocation density in the samples. Finally, a homogeneous distribution of $\theta'$ precipitates was found in the AA168 samples (Fig. \ref{fig:precipitados}c). The $\theta$ phase was not found in the AA168 alloy, in agreement previous first principles calculations \citep{WO01, LPL19} that have shown that $\theta'$ (and not $\theta$) is the stable phase at 180 $^o$C due to the vibrational entropic contribution.
 
\begin{figure}[h!]
	\centering
	\includegraphics[width=\textwidth]{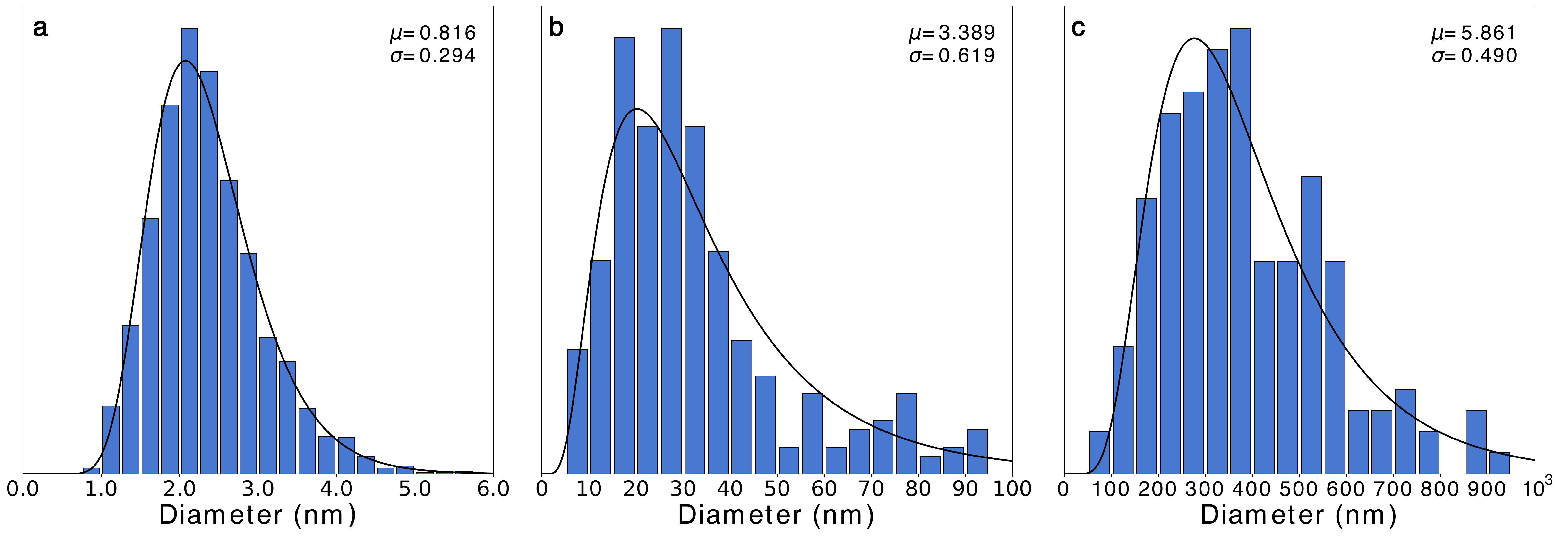}
	\caption{Distribution of the diameters of the precipitates for each aging condition. (a) NA alloy. (b)  AA30 alloy. (c) AA168 alloy. The mean $\mu$ and standard deviation $\sigma$ of the lognormal precipitate distributions are indicated in each plot.}
	\label{fig:histogram}
\end{figure}

 The diameter distributions of the precipitates for the three aging conditions are plotted in Fig. \ref{fig:histogram}. They followed a lognormal distribution and the corresponding parameters (mean $\mu$ and standard deviation $\sigma$) are indicated in each plot. The $\theta''$ and $\theta'$ precipitate volume fraction ($f$) was obtained from the precipitate number density ($n_p$), average precipitate diameter ($d$) and thickness ($t$) and foil thickness. It was not possible to measure the actual thickness of the GP zones because of the large foil thickness ($\approx$ 124 nm), as compared with the diameter of the GP zones \cite{Phillips1973, Karlik2004}. All the microstructural information about the precipitate distribution is  summarized in Table \ref{tab:precipitate-size} for each aging condition. In this table, the average and standard deviation of the diameter and thickness are those corresponding to a Gaussian distribution. 

\begin{table}[h!]
	\centering
	\caption{Average and standard deviation of the diameter, $d$, thickness, $t$,  number density, $n_p$ and volume fraction $f$ of the precipitates in the samples naturally aged (NA) and artificially aged at 180$^\circ$ during 30 hours (AA30) and 168 hours (AA168). }
	\begin{tabular}{lcccccc}
		\toprule
		Heat   & Precipitate & $d$  & $t$  &  $n_p$  & $f$ \\
		 treatment &  & (nm) & (nm) &   ($\mu$m$^{-3}$) & (\%) \\
		\midrule
NA	& GP & 2.4$ \pm $0.7 & --- &  4.3$\pm$0.5 $\times$ 10$^{6}$ & ---\\
AA30 &$\theta''$  &  36$ \pm $26 & 1.6$ \pm $0.2 &  9717$ \pm $6735 & 0.9$\pm $0.2 \\
AA168 & $\theta'$ & 393$ \pm $187 & 8.3$\pm$2.8 & 11$ \pm $2 & 1.0$ \pm $0.3 \\
		\bottomrule
	\end{tabular}%
	\label{tab:precipitate-size}%
\end{table}%

Finally, dislocations were observed in a $(1\bar{1}0)$ reflection in a $\langle110\rangle_{\alpha}$ zone axis prior to deformation and the initial average dislocation density $\rho \approx$ 2.2 10$^{13}$ m$^{-2}$ was determined from 10 micrographs in the NA alloy from the total dislocation length, the foil thickness and the total area analyzed.  Quantitative evaluations of the initial dislocation density  in the AA30 and AA168 alloys were not carried out but the TEM images showed a dislocation density similar to the one measured in the NA alloy.

\subsection{Analysis of the size effect}

\begin{figure}[h!]
	\centering
	\includegraphics[width=\textwidth]{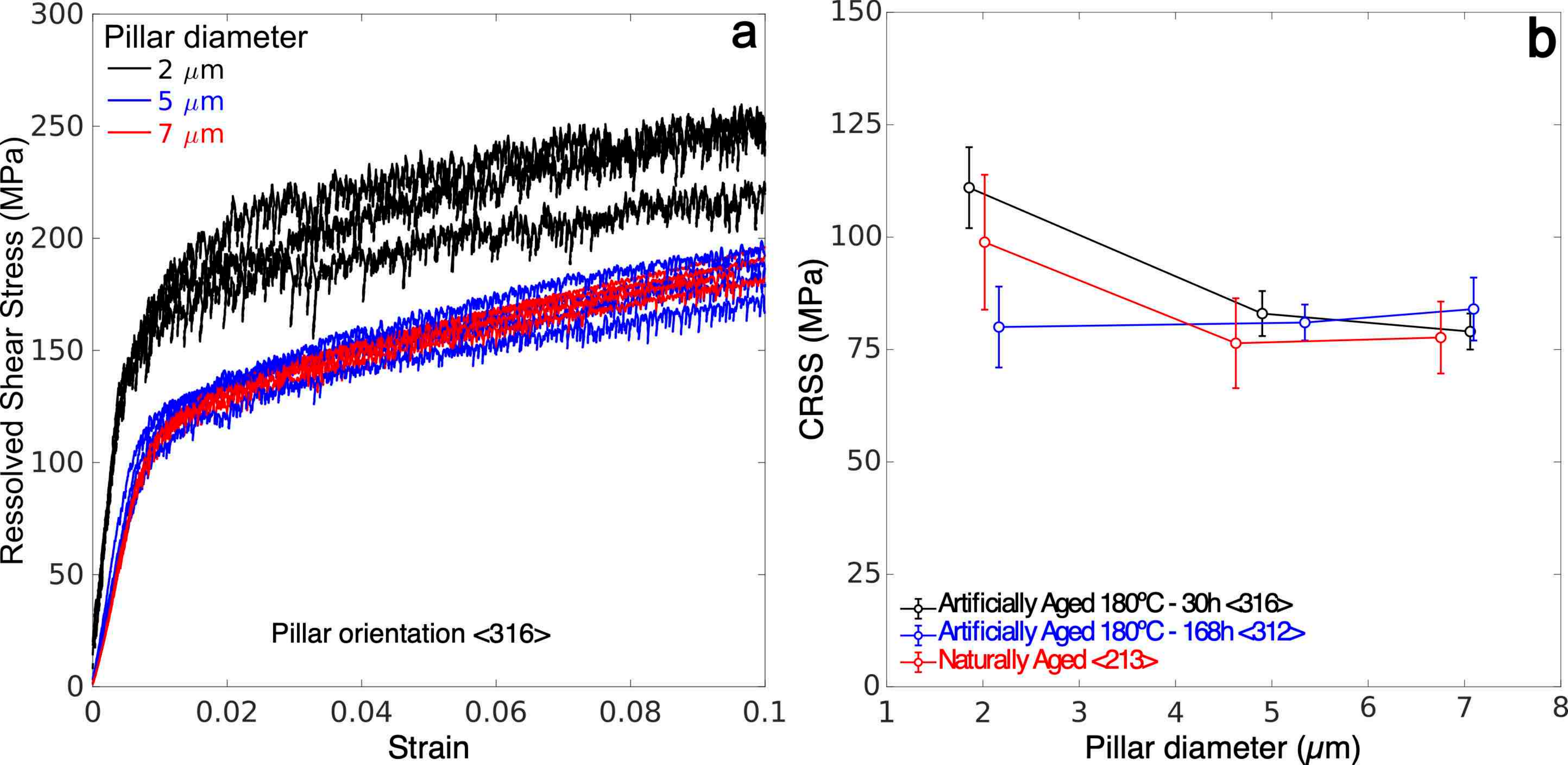}
	\caption{(a) Influence of the micropillar diameter on the RSS vs. strain curves of micropillars oriented for single slip in the sample aged at 180$^\circ$C during 168 hours. (b) CRSS as a function of the micropillar diameter in samples subjected to different aging treatments. The micropillar orientation is indicated in the legend.}
	\label{fig:SizeEffect}
\end{figure}

The yield stress measured in small samples, such as micropillars, may depend strongly on the dimensions of the micropillar \cite{Greer2011,Soler2012} and this size effect depends on the material. So, the first step in this investigation was to ascertain the influence of the micropillar diameter in the mechanical response. To this end, micropillars oriented for single slip with different diameter were carved from the same grain in each of the samples subjected to different heat treatments. The RSS  {\it vs.} strain curves in the case of the AA30 alloy are plotted in Fig. \ref{fig:SizeEffect}a. The noise in the curves of the smallest micropillars (2 $\mu$m in diameter) cannot hide the strong size effect of the type "smaller is stronger" found for this size. Nevertheless, this size effect has disappeared in the micropillars of 5 and 7 $\mu$m in diameter, which show the same values of the yield strength and of the strain hardening after yielding. Similar results were obtained in the case of NA samples while no size effect was found for micropillars with diameters in the range 2 to 7 $\mu$m in the case of the AA168 micropillars. The corresponding values of the CRSS as a function of the micropillar diameter are plotted in Fig. \ref{fig:SizeEffect}b.

Size effects in plasticity are associated to the ratio between the length scale that controls the movement of the dislocations (average dislocation distance, distance between obstacles) and the dimensions of the specimen. In the case of the precipitation-hardened alloys, the main obstacle to the dislocations are the precipitates and the average value of the minimum distance between them in the \{111\} slip plane, $\bar\lambda_p$ in the case of the disk-shaped precipitates parallel to the \{100\} planes of an FCC lattice is given by \cite{Nie2014} 

\begin{equation}
{\bar\lambda}_p= \frac{1.030}{\sqrt{n_p d}} - \frac{\pi}{8} d -1.061t
\end{equation}

\noindent where the precipitate diameter and thickness ($d$ and $t$) and number density, $n_p$, can be found in Table \ref{tab:precipitate-size} and the average distance between precipitates can be found in Table \ref{tab:distance}. In the case of the GP zones, their thickness was assumed to be equal to the diameter of the Cu atom (0.256 nm) \cite{Greenwood1997}. $\bar\lambda_p$ was at least one order of magnitude smaller than the micropillar diameter and, thus, size effects are expected to be negligible for micropillars $\ge$ 5 $\mu$m, in agreement with previous experimental observations in other precipitate-strengthened alloys \cite{Cruzado2015}. It should be noted, however, that a noticeable size effect appeared in the micropillars containing either GP zones or $\theta''$ precipitates when the micropillar diameter was 2 $\mu$m even though the precipitate spacing in both cases was more than two orders of magnitude smaller than the micropillar diameter. This size effect may appear in these cases because the dislocations can shear the precipitates  and leave the crystal through the lateral surfaces, limiting the density of mobile dislocation available to accommodate the imposed deformation. Shearing of GP zones is known to occur \cite{Byrne1961, Nie2014, EBM20} while $\theta''$ precipitates with a diameter below $\approx$ 30 nm are also sheared by dislocations, as it will be shown below.

\begin{table}[h!]
	\centering
	\caption{Average value of the minimum distance between precipitates, $\bar\lambda_p$, in the \{111\} slip plane for samples subjected to different heat treatments.}
	\begin{tabular}{lcccccc}
		\toprule
		Heat  treatment & Precipitate & $\bar\lambda_p$ (nm) \\
		\midrule
		NA	& GP & 8.9  \\
		AA30 & $\theta''$  &  39  \\
		AA168  & $\theta'$ & 332 \\
		\bottomrule
	\end{tabular}%
	\label{tab:distance}%
\end{table}%

\subsection{Effect of precipitate type on strengthening}

The effect of the precipitate type on the strength of the Al-Cu alloys was analyzed by comparing the RSS {\it vs.} plastic strain curves in micropillar of square cross-section of  5 x 5 $\mu$m$^2$ oriented for single slip. The plastic strain (obtained as the total strain minus the elastic contribution given by the stress divided the elastic modulus of the micropillar measured in the final elastic unloading) was used to make the comparison instead of the total strain. In this way, the experimental uncertainties associated with the initial slope of the curves (that can easily affected by the roughness of the pillar surface and the misalignment between the flat punch and the top of the pillar) were eliminated. 

\begin{figure}[h!]
	\centering
	\includegraphics[height=8cm]{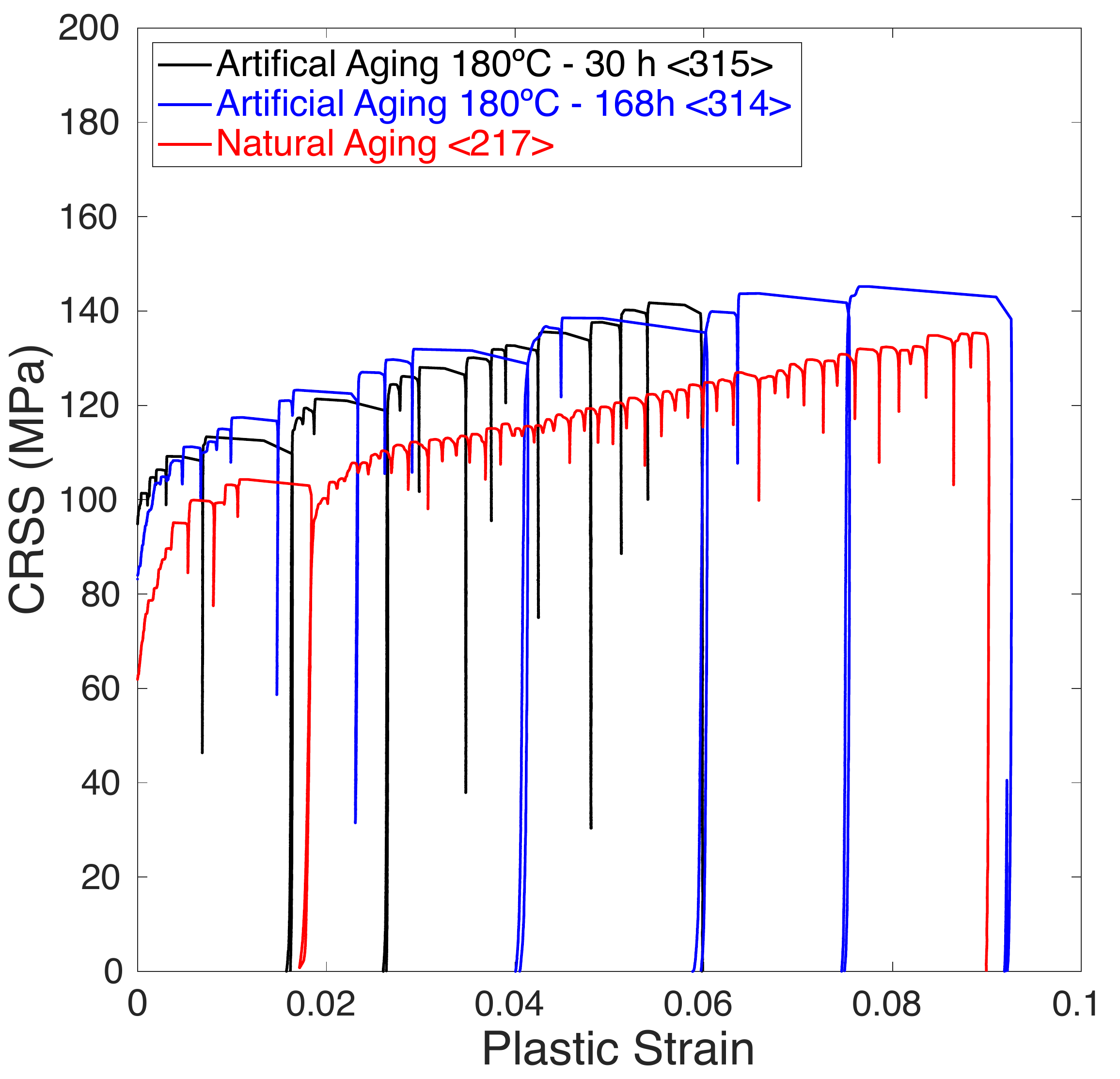}
	\caption{Representative RSS {\it vs.} plastic strain curves of micropillars oriented for single slip in the alloy subjected to different aging conditions. The orientation of each micropillar is indicated in the legend.}
	\label{fig:Precipitateeffect}
\end{figure}

The experimental RSS {\it vs.} plastic strain curves obtained from 3 representative  micropillars oriented for single slip of the NA, AA30 and AA168 alloy are plotted in Fig. \ref{fig:Precipitateeffect}. All the curves (but particularly those of the artificially aged alloys) shown strain bursts --associated with a sudden drop in the stress-- which were caused by the avalanches  of dislocations that escape from micropillar through the lateral surfaces. The reduction of the stress in these cases is an artefact due to the limitations of the PID control system to follow the sudden deformation of the micropillar.

The average CRSS (understood as the RSS at a plastic strain of 0.02\%) of each alloy is depicted in Table \ref{tab:CRSS}. The maximum value was obtained in the alloy aged at 180$^\circ$C during 30 hours, followed closely by alloy aged at the same temperature during 168 hours, while the minimum CRSS was measured in the alloy aged at ambient temperature. An initial evaluation of these results for the AA30 and AA168 alloys can be performed assuming that the main contributions to the initial CRSS are the forest, solution  and precipitate hardening as well as the interaction of the dislocations with the stress field around the precipitates due to the lattice mismatch between the matrix and the precipitate and that they can be added. The forest hardening contribution can be estimated from the Taylor model \cite{T34} according to 

\begin{equation}\label{Taylor}
\tau_{f}= \alpha \mu b\sqrt{\rho},
\end{equation}

\noindent where $\mu$ = 25 GPa stands for the shear modulus of Al, $b$ = 0.286 nm the Burgers vector \cite{RHL19}, $\rho$ the initial dislocation density (2.2 10$^{13}$ m$^{-2}$) and $\alpha$ = 0.7  is a constant that accounts for the different dislocation interactions \citep{KM03}. 
The influence of the Cu atoms in solid solution to the CRSS, $\tau_{ss}$ can be determined as \cite{ZS08, Alex2018},

\begin{equation}\label{eq:ss}
	\tau_{ss}=Hx_{cu}^n
\end{equation}

\noindent where $x_{cu}$ is the weight fraction of Cu dissolved in the Al matrix (which can be determined for each treatment by subtracting the Cu content in the precipitates from total weight fraction of 0.04), and $H$ and $n$ are two constants whose values in Al-Cu alloys are $H$ = 7.2 MPa and $n$ = 1 \cite{ZS08, Alex2018}. 

The strengthening provided by disk-shape precipitates parallel to the \{100\} planes of the FCC lattice due to the formation of Orowan loops, $\tau_p$ was estimated by Nie and Muddle a screw dislocation as \cite{Nie2008},

\begin{equation}\label{eq:p}
	\tau_{p} = \frac{\mu b}{2\pi\sqrt{1-\nu}}\Bigg[\frac{1}{0.931\sqrt{\frac{0.306 \pi d t}{f}}-\frac{\pi d}{8}-1.061t}\Bigg] \ln \bigg(\frac{1.225t}{b}\bigg)
\end{equation}

\noindent where  $\nu$ = 0.33 stands for the Poisson ratio of the Al matrix. Finally, the contribution of the stress field around the precipitate to the CRSS, $\tau_S$ (31 MPa), has been calculated by means of discrete dislocation dynamics simulations for the AA168 alloy \cite{SBE20}  and is included in Table \ref{tab:CRSS}. This large contribution to the strengthening of $\theta '$ precipitates has been recently traced to the stress-free transformation strain associated with the nucleation and growth of these precipitates \citep{SBE20}. As shown experimentally \citep{DW83}, the nucleation of the $\theta '$ disks parallel to the \{100\} planes of the FCC Al lattice can be described by an invariant plane strain deformation of the lattice which has a small volumetric component  and a large shear one (33\%) \citep{Nie2014}. The interaction of the dislocation of the stress-field created by the shear strain associated to the precipitate formation increases dramatically the CRSS necessary to overcome the precipitates \citep{SBE20}. In comparison, the stress field around the  $\theta ''$ precipitates is much smaller \citep{LPL19} and its contribution to the strengthening is negligible.

The estimated contributions to the CRSS of the forest, solution, precipitation  and stress field  around the precipitates are included in Table \ref{tab:CRSS}. $\tau_{ss}$ is similar for all aging conditions but the contribution of precipitation hardening for the AA30 alloy is higher than for the AA168 alloy. Nevertheless, the contribution of the stresses around the precipitates in the AA168 alloy is much higher than in the AA30 alloy. Overall, the combined contributions of forest, solution and precipitation hardening and of the stresses around the precipitates  are very close to the experimental value of the CRSS in the AA30 and AA168 alloys. The contributions of forest and solution hardening to the NA Alloy were included Table \ref{tab:CRSS} for the sake of completion, as it is known that GP zones are sheared by dislocations and the Orowan mechanisms is not applicable to predict the CRSS of these materials.

\begin{table}[h!]
	\centering
	\caption{Experimental CRSS of the Al-Cu alloys subjected to different heat treatments and estimations according to eqs.  \eqref{Taylor}, \eqref{eq:ss}, \eqref{eq:p}  and  $\tau_S$ \cite{SBE20} for the AA30 and AA168 alloys.}
	\begin{tabular}{lcccccc}
		\toprule
		Heat  treatment &  CRSS  & $\tau_f$ & $\tau_{ss}$ & $\tau_p$ & $\tau_{S}$ & $\tau_f + \tau_{ss} + \tau_p+\tau_{S}$ \\
	 &  (MPa) & (MPa) &  (MPa) &  (MPa) &  (MPa) & (MPa)\\
		\midrule
		NA	& 61 $\pm$ 4 &23.5   &  -- & -- &  -- & --\\
		AA30 & 91 $\pm$ 5  & 23.5  & 25.7 & 48.7 &  -- & 97.9  \\
		AA168 & 80 $\pm$ 7 & 23.5  & 25.2 & 14.3 & 31 & 94 \\
		\bottomrule
	\end{tabular}%
	\label{tab:CRSS}%
\end{table}%

\begin{figure}[h!]
	\centering
	\includegraphics[width=\textwidth]{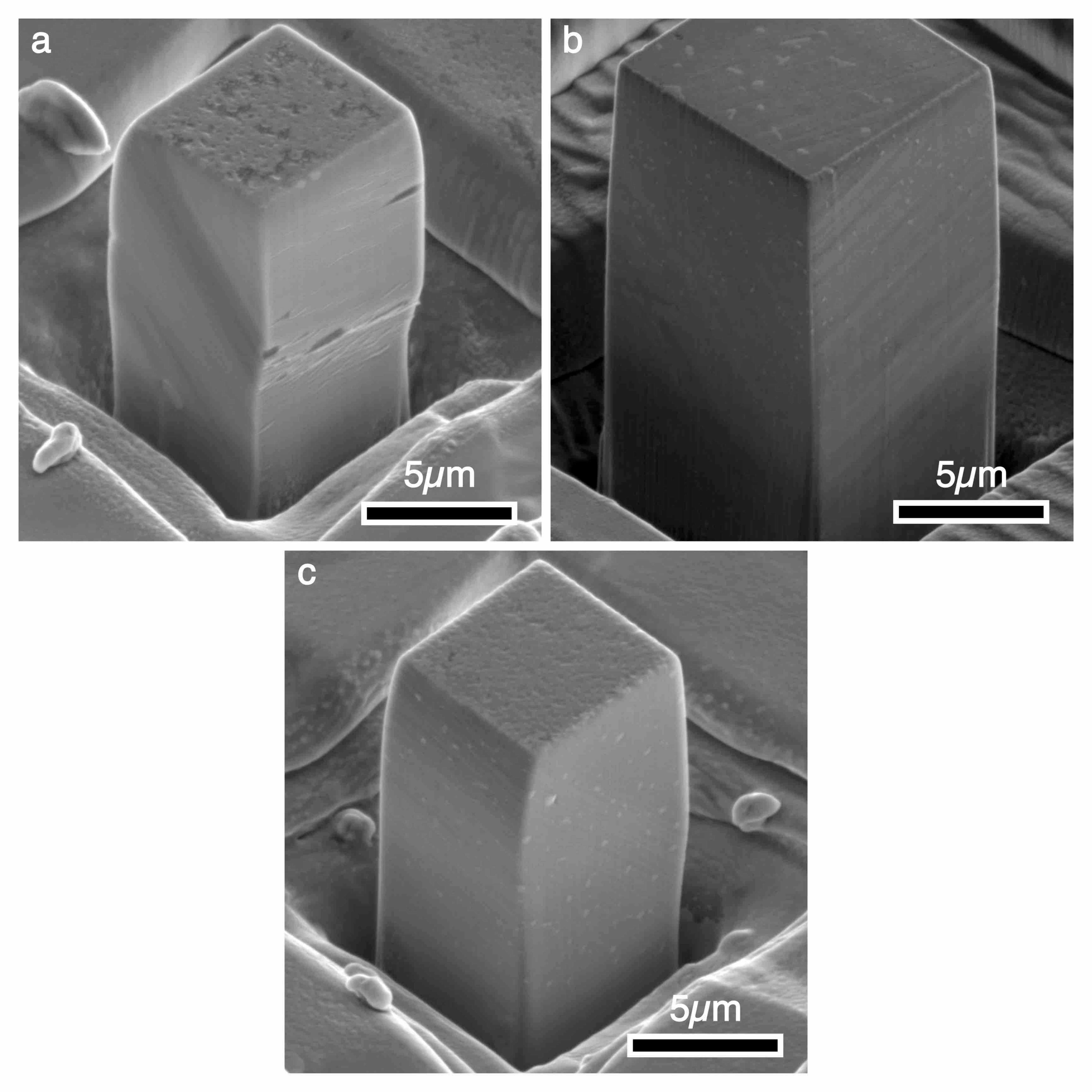}
	\caption{Scanning electron micrographs of micropillars deformed in single slip. (a) Naturally aged, showing the localization of deformation along slip bands. (b) Artificially aged during 30 hours 180$^{\circ}$C. (c) Artificially aged during 168 hours 180$^{\circ}$C. The artificially aged micropillars showed homogeneous deformation along the whole micropillars, as indicated by the slip traces.}
	\label{fig:Pilares_def_SS}
\end{figure}

The engineering RSS-plastic strain curves corresponding to the NA alloy in Fig. \ref{fig:Precipitateeffect} showed an initial parabolic hardening, which ended at a plastic strain of around 2\% with the formation of a strain burst. This strain burst was associated with the localization  of the deformation along the favourably oriented slip plane starting from the top of the micropillar, as shown in Fig. \ref{fig:Pilares_def_SS}a. The hardening in the RSS-plastic strain curve after this point is an artifact due to the increase of contact surface between the flat punch and the micropillar as a result of the slip band. Another distinct feature of the RSS-plastic strain curves of the NA alloy is the presence of serrations all along the curves due to dynamic strain aging, which is known to develop at ambient temperature in Al-4 wt. \%Cu alloys deformed at strain rates in the range 10$^{-5}$ to 10$^{-3}$ s$^{-1}$ \cite{JZJ05, JZC07}. This result is compatible with the hardening provided by GP zones that are shearable by dislocations. 

In the case of the artificially aged alloys, the engineering RSS-plastic strain curves for the micropillars oriented for single slip showed linear hardening after the elasto-plastic transition and the hardening rate was equivalent for both aging conditions. An increase of the hardening on the AA168 pillars was expected due to the forest hardening provided by the dislocation pilling up on the precipitates, however the higher precipitate number of the $\theta''$ present on the AA30, about three orders of magnitude higher than $\theta '$ on AA168 (Table \ref{tab:precipitate-size}), can act as artificial pinning points for the dislocation and contribute to the increase in the forest hardening, thus increasing the hardening rate. Some serrations were occasionally observed but it is obvious that dynamic strain aging was not one of the dominant deformations mechanisms. The deformed micropillars (Figs. \ref{fig:Pilares_def_SS}b and c) did not show any evidence of localization of the deformation along slip bands and plastic deformation occurred along the length of the micropillar, as shown by the slip traces on the micropillar surface. It should be noted that the slip traces were more difficult to find in the artificially aged micropillars because crystals oriented for single slip may be forced to deform via multiple slip due to the presence of presence of precipitates that are impenetrable to dislocations \cite{Sehitoglu2005, Byrne1961, Russell1970, Price1964}.  

\begin{figure}[h!]
	\centering
	\includegraphics[width=\textwidth]{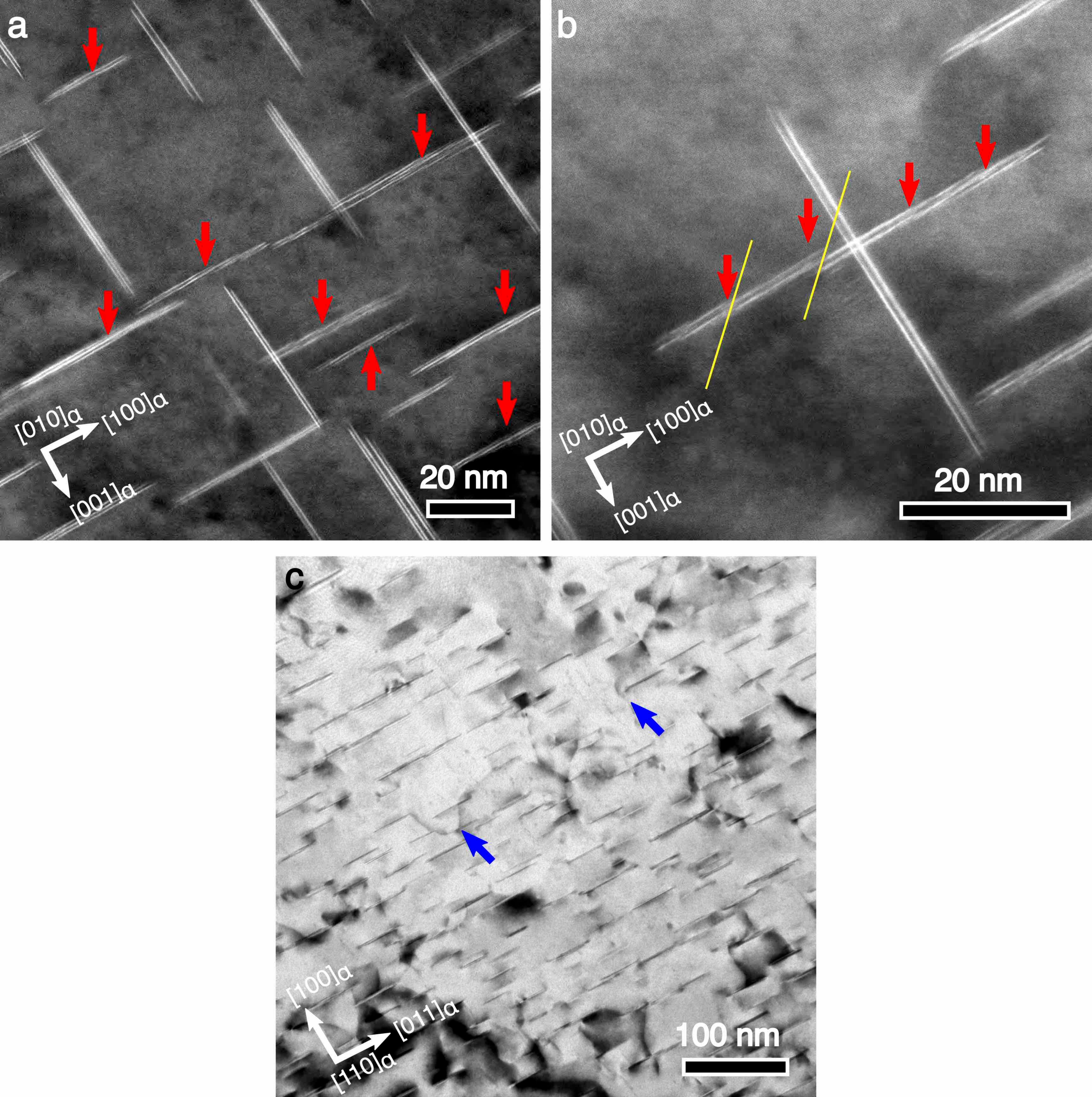}
	\caption{(a) and (b) High-angle annular dark-field/transmission electron micrographs of lamella parallel to [010]$_{\alpha}$ zone axis extracted from the alloy artificially aged during 30 hours. Small  $\theta''$ precipitates (diameter around 40 nm) were sheared by dislocations. The red arrows indicate the planes in which the layers of Cu atoms have been disrupted by the shearing of the dislocation. The direction of the Burgers vector ([101]) is indicated by the yellow lines. (c) High-resolution transmission electron micrograph showing the formation of Orowan loops (blue arrows) between large $\theta''$ precipitates (diameter above  $\approx$ 50 nm). The lamella was parallel to [010]$_{\alpha}$ zone axis.}
	\label{fig:Precipitados cortados}
\end{figure}

It was assumed that the $\theta'$ precipitates blocked the dislocation slip in the alloy artificially aged during 180 hours \cite{DaCostaTeixeira2008, Nie2008, Sehitoglu2005, Kaira2019} but the mechanisms of dislocation/precipitate interaction in the AA30 alloy were studied by TEM in thin foils extracted from the deformed micropillars. Small $\theta''$ precipitates with a diameter below 40 nm and including two layers of Cu atoms were sheared by dislocations, as indicated by the red arrows in Figs. \ref{fig:Precipitados cortados}a and b. The shearing direction coincided with the direction of the Burgers vector [101] in the (111) slip plane, that is indicated by yellow lines. Nevertheless, dislocations were not able to shear larger $\theta''$ precipitates, with diameters above  $\approx$ 50 nm, and formed Orowan loops to overcome these precipitates (Fig. \ref{fig:Precipitados cortados}c). Approximately, 25\% of the $\theta''$ precipitates in the AA30 alloy had a diameter above 50 nm (Fig. \ref{fig:histogram}b). Their average  diameter was 90 nm  and the average distance between them was also approximately 90 nm, assuming that they were homogeneously distributed. Their presence contributed to limit strain localization in the AA30 alloy and also to increase the strain hardening rate, which was equivalent to the one in the AA168 alloy. Thus, the maximum strength was obtained with a fine and homogeneous dispersion of $\theta''$ precipitates whose average dimensions were close to the transition from precipitate shearing to the formation of Orowan loops, in agreement with the accepted theories for optimum precipitate strengthening \cite{Nie2014, Polmear2017}.

Finally the blue curve in Fig. \ref{fig:Precipitateeffect} corresponds to the AA168 alloy which contains $\theta'$ precipitates. Dislocations cannot shear these precipitates and they are by-passed by the formation of Orowan loops \cite{Nie2014,Nie2008,Santos-Guemes2018}. Although the diameter and spacing of the $\theta '$ precipitates in this sample were much larger that those found in the sample artificially aged during 30 hours, the CRSS and the hardening rate were surprisingly very similar for both samples. The excellent strengthening capability of $\theta '$ precipitates has been recently traced to the contribution of the stress-free transformation strain associated with the nucleation and growth of these precipitates to the CRSS \citep{SBE20}. 

\subsubsection{Effect of micropillar orientation}

\begin{figure}[!]
\centering
         \includegraphics[width=0.85\textwidth]{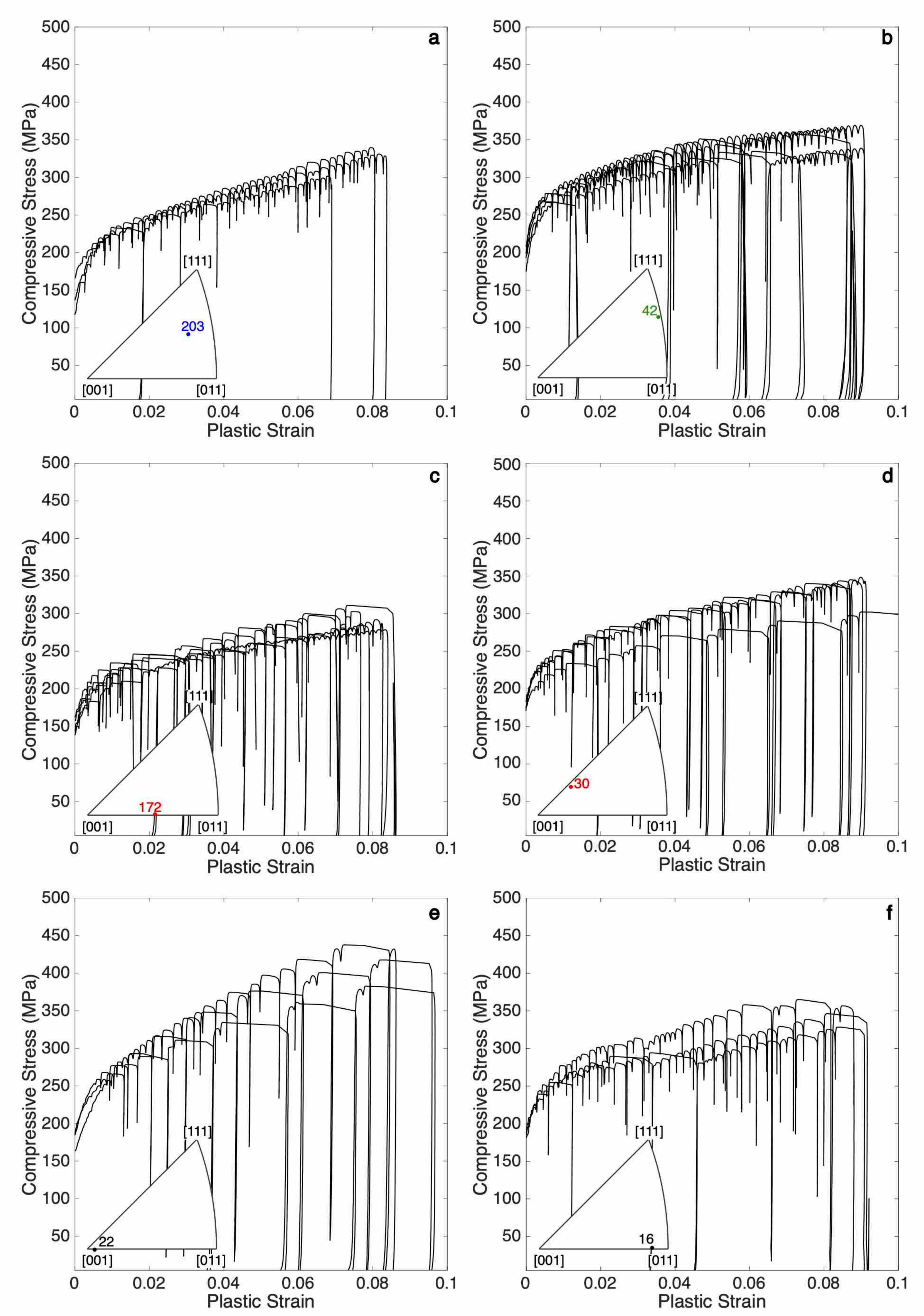}
         \caption{Engineering compressive stress {\it vs.} plastic strain curves of micropillars with different orientation of the naturally aged alloy. (a) Single slip. (b) Double coplanar slip. (c) Double non-coplanar slip. (d) Multiple slip. (e) Multiple slip. (f) Multiple slip. The orientation of each micropillar is indicated in the inverse pole figures.}
        \label{fig:OrientationNA}
\end{figure}

\begin{figure}[h!]
	\centering
	\includegraphics[width=0.5\textwidth]{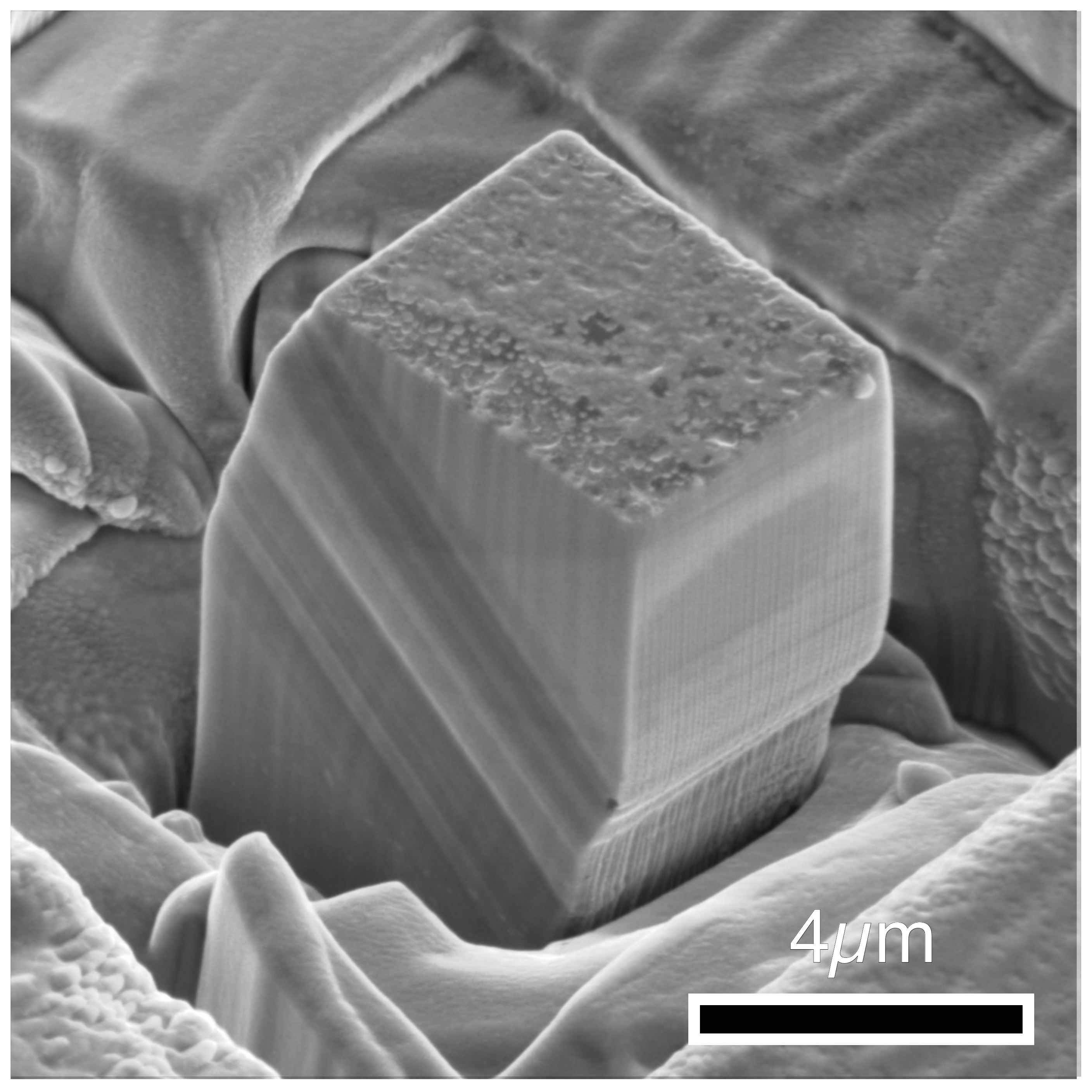}
	\caption{Scanning electron micrograph of a naturally aged micropillar oriented for non-coplanar double slip, showing the localization of deformation along slip bands.}
	\label{fig:PilarDSNC}
\end{figure}

Compression tests were carried out in micropillars carved from grains with different orientations to promote single slip, double slip (coplanar and non-coplanar) and multiple slip in materials in the three aging conditions. Three or four micropillars were tested for each grain orientation and the grain number, crystallographic orientation and maximum Schmid factor (SF) are shown in Table \ref{tab:orientations}. The engineering compressive stress {\it vs.} plastic strain curves are plotted in Figs. \ref{fig:OrientationNA} for the NA alloy. All the curves showed the presence of strain bursts --associated with a sudden drop in the stress-- but the frequency of these events was higher in micropillars oriented for double and multiple slip, in comparison with those oriented for single slip. In addition, serrations due to dynamic strain aging were found in the micropillars oriented for single and double coplanar slip (Figs. \ref{fig:OrientationNA}a and b) but not in the micropillars oriented for non-coplanar double slip or multiple slip. These differences may be due to the dominance of latent hardening (induced by the interaction of dislocations gliding on different slip systems) with respect to dynamic strain aging in the latter micropillars. Finally, it should be noted that strain localization was found in all the NA micropillars oriented for single or double slip (Fig. \ref{fig:Pilares_def_SS}a and Fig. \ref{fig:PilarDSNC}) and in some of the micropillars oriented for multiple slip. Thus, the strong strain hardening observed in the stress- plastic curves in Fig. \ref{fig:OrientationNA} was partially an artefact and they could not be used to calibrate the crystal plasticity model.

\begin{table}[h!]
\centering
\caption{Grain number, crystallographic orientations and maximum Schmid Factors (SF) for each micropillar tested in compression}
\begin{tabular}{ccccccccc}
\toprule
\multicolumn{3}{p{10em}}{Naturally aged}        & \multicolumn{3}{p{10em}}{Artificially aged 30 h} & \multicolumn{3}{p{10em}}{Artificially aged 168 h}\\
Grain           &   Orient.   & SF &   Grain    &   Orient.  & SF &   Grain    &   Orient.  &   SF   \\
\midrule
203    & [213]  & 0.46       & 53  & [315] & 0.50        & 33  & [314] & 0.47 \\
42   & [212]  & 0.39       & 46  & [122] & 0.41       & 30  & [313] & 0.43 \\
172  & [102]  & 0.49       & 2   & [114] & 0.46       & 14  & [116] & 0.46 \\
22   & [001]  & 0.43       & 43  & [111] & 0.30       & 19  & [101] & 0.43 \\
30   & [116]  & 0.46       & 35  & [001] & 0.44       & 42  & [102] & 0.50  \\
16   & [101]  & 0.45       &   &  &       & 17  & [001] & 0.42 \\
\bottomrule
\end{tabular}
\label{tab:orientations}
\end{table}

\begin{figure}[!]\centering
         \includegraphics[width=0.85\textwidth]{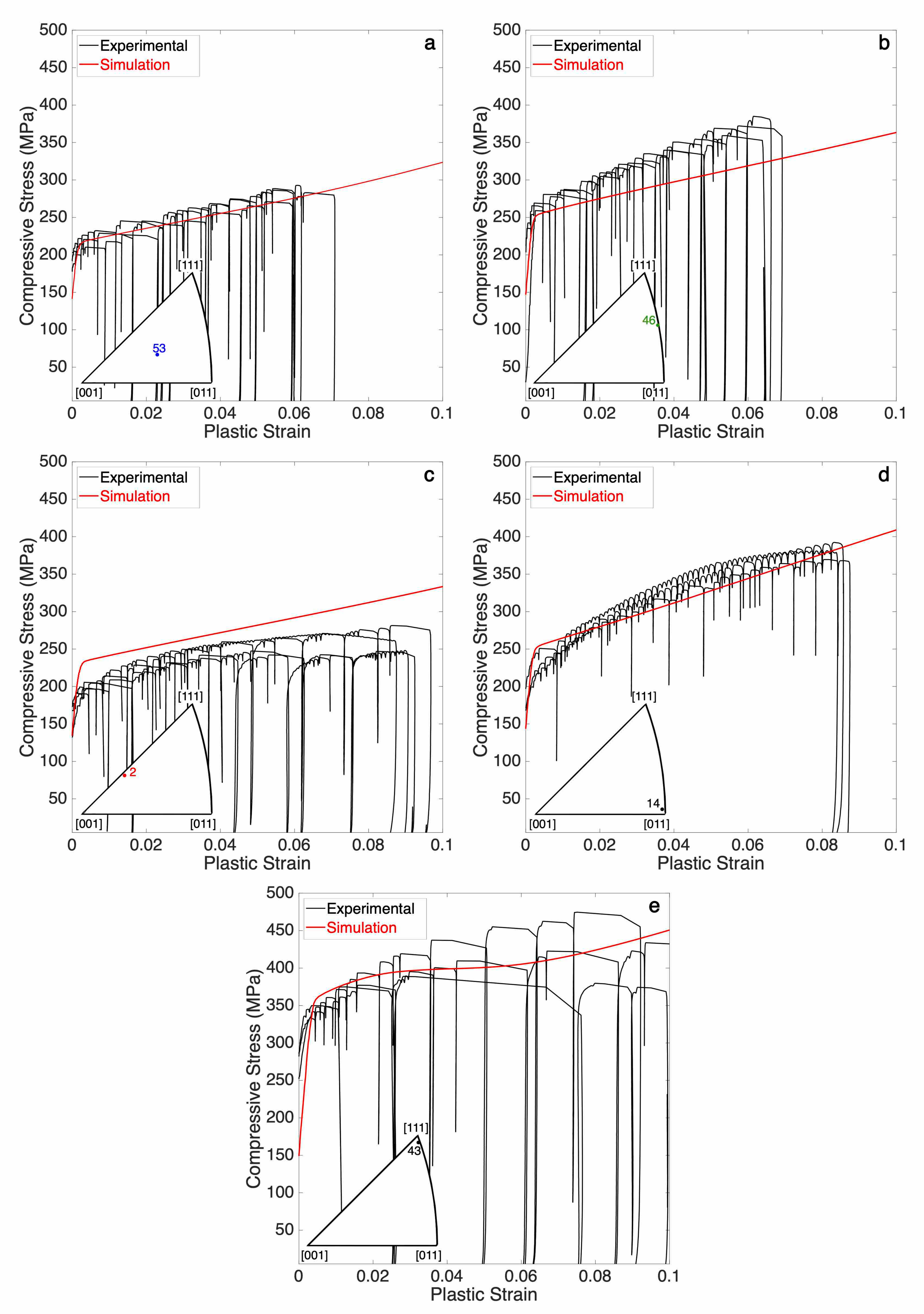}
         \caption{Engineering compressive stress {\it vs.} plastic strain curves of micropillars with different orientation of the artificially aged alloy at 180 $^\circ$C during 30 hours. (a) Single slip. (b) Double coplanar slip. (c) Double non-coplanar slip. (d) Multiple slip. (e) Multiple slip. The orientation of each micropillar is indicated in the inverse pole figures. The red curves stand for the results of the simulation of the compression tests using the crystal plasticity model.}
        \label{fig:OrientationAA30}
\end{figure}

\begin{figure}[!]\centering
         \includegraphics[width=\textwidth]{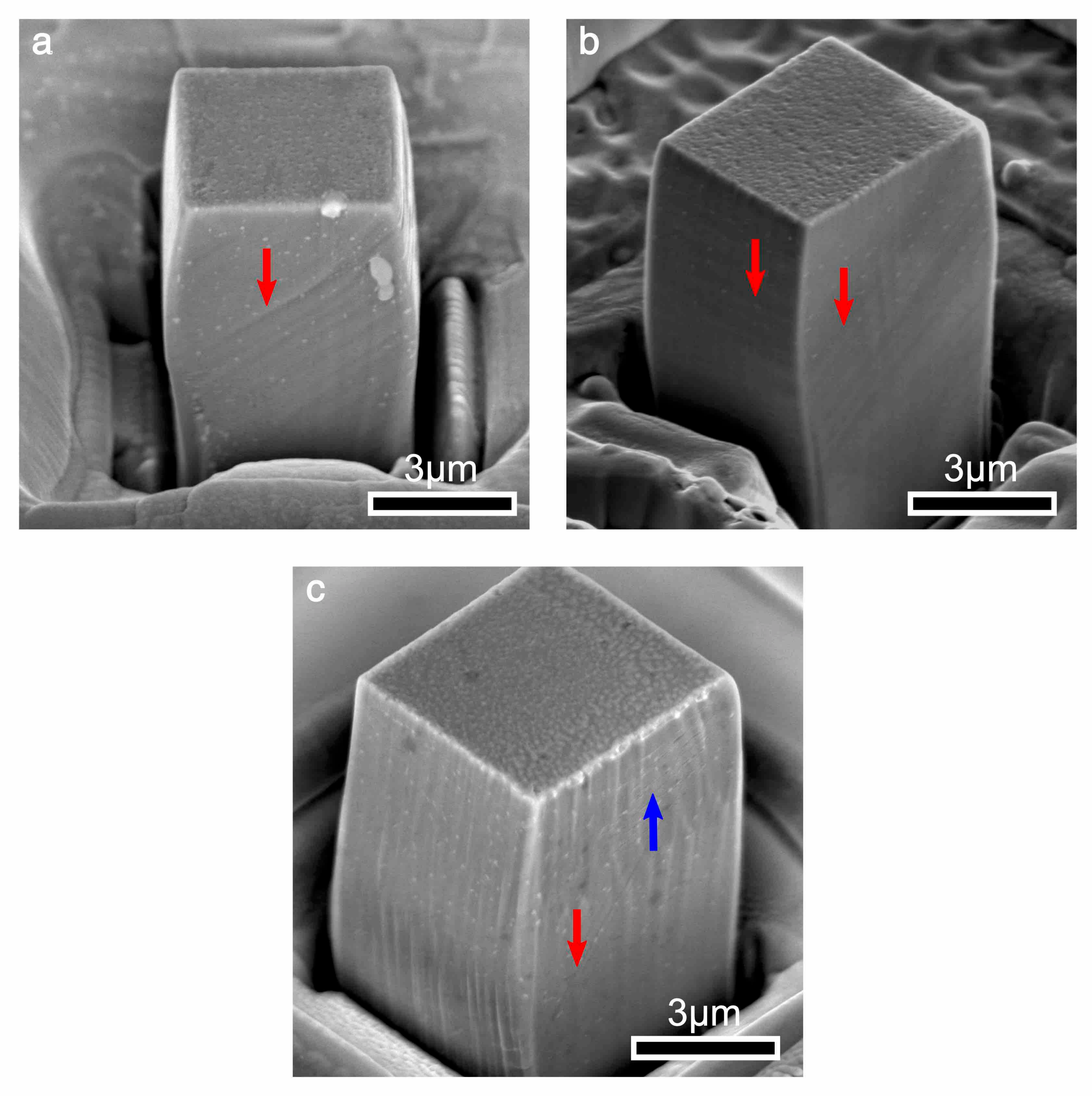}
         \caption{SEM micrographs of deformed micropillars of the artificially aged alloy at 180 $^\circ$C during 30 hours. (a) Double  coplanar slip, showing some localization of deformation along slip bands. (b) Double non-coplanar slip. (c) Multiple slip. Slip bands have been marked with red arrows, blue arrows indicate wavy slip traces.}
                 \label{fig:DefAA30}
\end{figure}

\begin{figure}[!]\centering
         \includegraphics[width=0.85\textwidth]{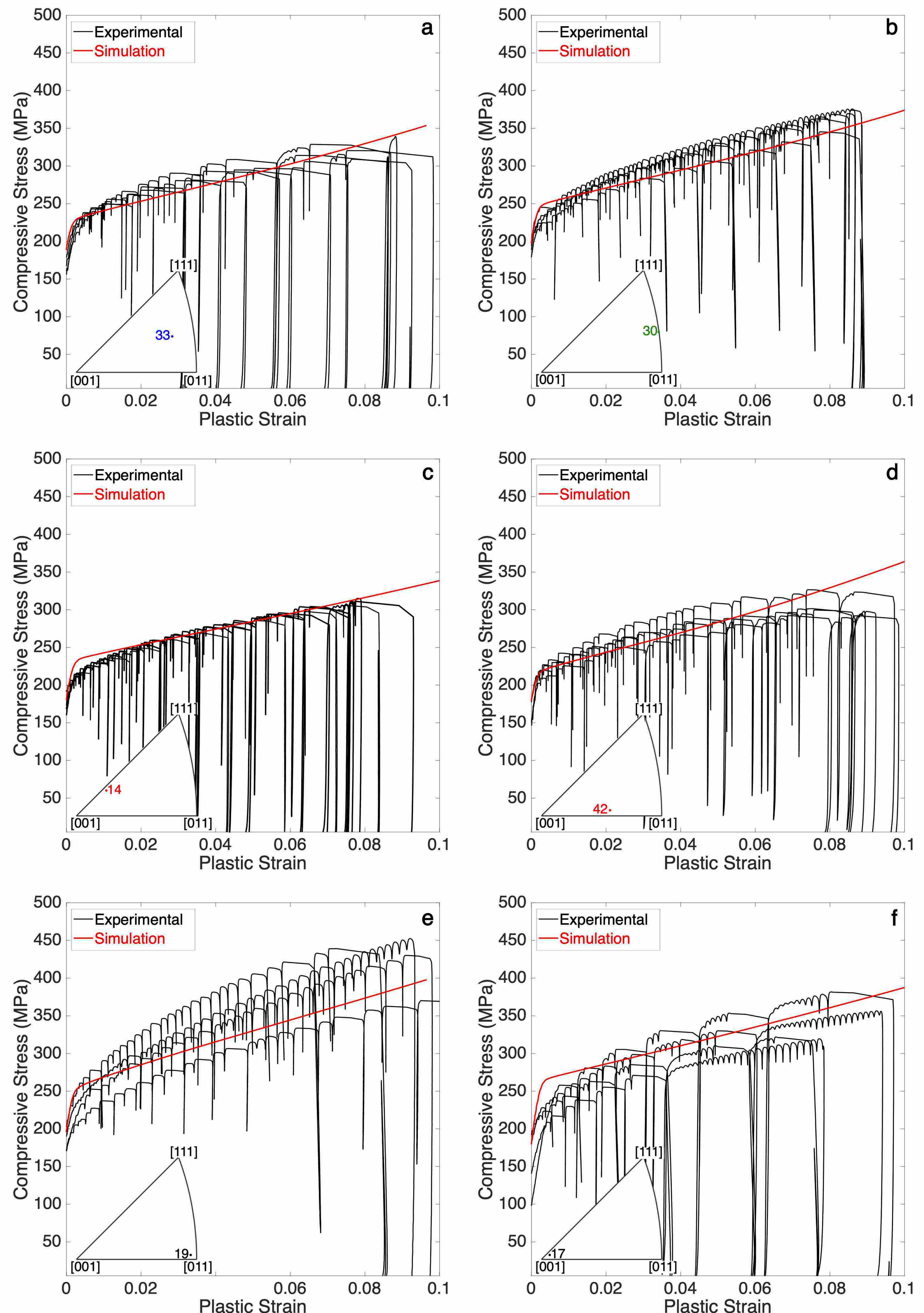}
         \caption{Engineering compressive stress {\it vs.} plastic strain curves of micropillars with different orientation of the artificially aged alloy at 180 $^\circ$C during 168 hours. (a) Single slip. (b) Double coplanar slip. (c) Double non-coplanar slip. (d) Double non-coplanar slip. (e) Multiple slip. (f) Multiple slip. The orientation of each micropillar is indicated in the inverse pole figures. The red curves stand for the results of the simulation of the compression tests using the crystal plasticity model.}
        \label{fig:OrientationAA168}
\end{figure}

\begin{figure}[!]\centering
         \includegraphics[width=\textwidth]{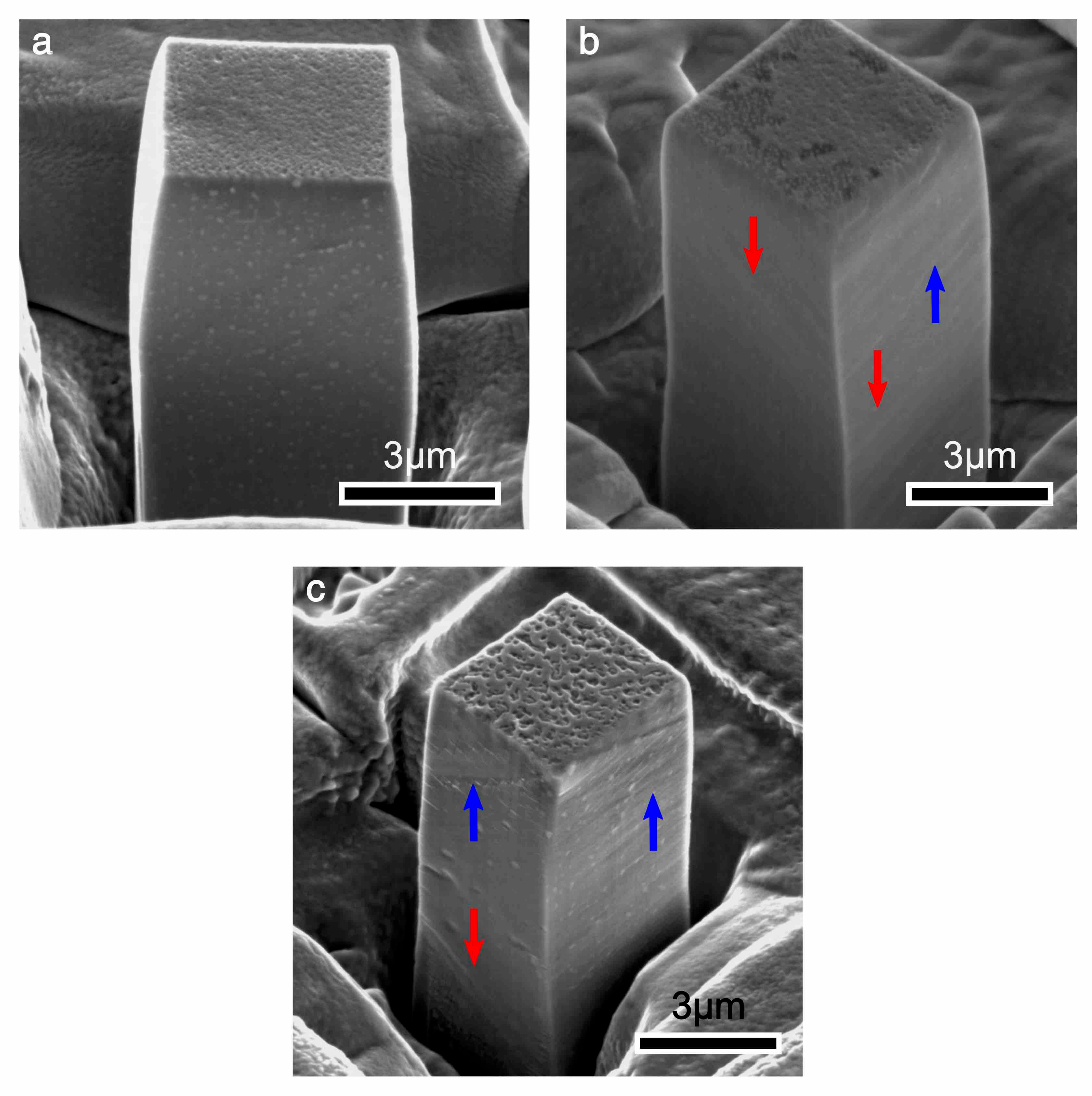}
         \caption{SEM micrographs of deformed micropillars of the artificially aged alloy at 180 $^\circ$C during 168 hours. (a) Double  coplanar slip, showing some localization of deformation along slip bands. (b) Double non-coplanar slip. (c) Multiple slip. Slip bands have been marked with red arrows, blue arrows indicate wavy slip traces.}
        \label{fig:DefAA168}
\end{figure}

The engineering compressive stress {\it vs.} plastic strain curves for the AA30 micropillars with different orientations are plotted in Fig. \ref{fig:OrientationAA30} while three representative SEM micrographs of these micropillars deformed under coplanar and non-coplanar double slip as well as multiple slip are shown in Figs. \ref{fig:DefAA30} (a), (b) and (c), respectively. Similarly, the corresponding curves for the AA167 micropillars with different orientations are plotted in Fig. \ref{fig:OrientationAA168} while three representative SEM micrographs of these micropillars deformed under coplanar and non-coplanar double slip as well as multiple slip are shown in Figs. \ref{fig:DefAA168} (a), (b) and (c), respectively.
The experimental scatter was very limited for most orientations, indicating the micropillar compression is a reliable technique to determine the behavior of single crystals in the absence of strain localization. The differences in the initial yield stress among the curves can be readily explained by the differences in the SF among orientations, while the strain hardening afterwards is related to the interaction among dislocations gliding in different systems. Nevertheless, as indicated in \citep{Sehitoglu2005}, the plastic flow anisotropy between orientations is reduced by the presence of precipitates and the smallest anisotropy was found in the AA168 micropillars. The minimum hardening was always found in the micropillars oriented for single slip and double, non-coplanar slip because the interactions among dislocations were minimum in these configurations. The hardening rate increased in the case of double, coplanar slip and was maximum for the micropillars oriented for multiple slip. Based on these results, a phenomenological crystal plasticity model was used to predict the plastic response of the alloys under different aging conditions, as shown in the following section. 

The micropillar compression tests in crystals oriented for single and double slip clearly showed that deformation tends to localize in crystallographic planes in the NA alloy because the GP zones can be sheared by dislocations (Figs. \ref{fig:Pilares_def_SS} (a) and \ref{fig:PilarDSNC}). This is not the case in the AA30 and AA168 alloys, where the presence of precipitates that cannot be sheared by dislocations forces to deform via multiple slip through cross-slip. As a result, the deformed micropillars of the AA30 and AA168 alloys tend to have a "barrel" shape (Figs. \ref{fig:DefAA30} and \ref{fig:DefAA168}), which is indicative of homogeneous deformation along the micropillar \citep{Sehitoglu2005} while those of the NA alloy showed the localization of deformation in a few slips bands along the micropillar. Nevertheless, straight slip traces (marked with red arrows) were more clearly observed in the AA30 alloy, particularly for coplanar and non-coplanar double slip, Figs. \ref{fig:DefAA30}(a) and (b), respectively.  This observation is compatible with shearing of some of the $\theta''$ precipitates in this aging condition. On the contrary, most of the slip traces observed in the AA168 alloy deformed under double or multiple slip (marked by blue arrows) are wavy, and the straight slip traces are very faint (Fig. \ref{fig:DefAA168}). These results indicate that the presence of the impenetrable $\theta'$ precipitates promotes multiple slip through the activation of cross-slip. It should be finally noted that some vertical lines on the deformed micropillars in Figs. Figs. \ref{fig:DefAA30} and \ref{fig:DefAA168} are not slip traces but marks left by the FIB beam during milling.

\section{Crystal plasticity modelling}\label{S:3}
\subsection{Crystal plasticity model}

The mechanical behavior of the Al-Cu single crystals artificially aged at 180$^\circ$C was based in a phenomenological crystal plasticity model that was successfully used to predict the mechanical response of precipitation-hardened Ni-based superalloys \cite{Cruzado2015}. Nevertheless, this phenomenological model is not appropriate to simulate the behavior of the naturally-aged micropillars, which showed a very marked strain localization during deformation.

The model is based on the multiplicative decomposition of the deformation gradient, $\textbf{F}$ into the elastic ($\textbf{F}^e$)  and plastic ($\textbf{F}^p$) components, 

\begin{equation}\label{eq-defGradient}
	\textbf{F}=\textbf{F}^e\textbf{F}^p.
\end{equation}

The plastic velocity gradient, $\textbf{L}^p=\dot{\textbf{F}}^p\textbf{F}^{-1}$, is obtained as the contribution of the 
 shear rate, $\dot{\gamma}^{\alpha}$ of each slip system $\alpha$ according to \cite{Segurado2018}
 
\begin{equation}
\textbf{L}^p=\sum^{N_{slip}}_{\alpha=1} \dot{\gamma}^{\alpha}(\textbf{s}^{\alpha}_0\otimes\textbf{m}^{\alpha}_0)
\end{equation}

\noindent where $\textbf{s}^{\alpha}_0$ and $\textbf{m}^{\alpha}_0$ stand, respectively, for unit vectors in the slip direction and normal to the slip plane in the reference configuration and $N_{slip}$ is the number of slip systems.

The crystal is assumed to behave as an elasto-viscoplastic solid in which the plastic slip rate of each slip system follows a power-law dependency,

\begin{equation}
\dot{\gamma}^{\alpha}=\dot{\gamma}_{0}\left( \frac{|\tau^{\alpha}|}{\tau^{\alpha}_c}\right) ^{1/m}sgn(\tau_{\alpha})
\end{equation}

\noindent where $\dot{\gamma}_{0}$ is a reference shear strain rate, $\tau^{\alpha}_c$ the CRSS on the slip system, $m$ the rate-sensitivity exponent and $\tau^{\alpha}$ stands for the resolved shear stress on the slip system.

The evolution of the CRSS on a given system $\alpha$, is expressed as \cite{Cruzado2015}

\begin{equation}
\dot{\tau}^{\alpha}_c=\sum_{\beta}=q_{\alpha\beta}h|\dot{\gamma}^{\beta}|
\end{equation}

\noindent where $\beta$ stands for any slip system. $q_{\alpha\beta}$ are the strain hardening coefficients due to self ($\alpha=\beta$) and latent hardening ($\alpha\neq\beta$), and $h$ is the self hardening modulus. The evolution of the self hardening was described using the Voce hardening model

\begin{equation}\label{eq-voceHardening}
h(\gamma_{a})=h_s+\left[h_0 -h_s \frac{h_0h_s\gamma_a}{\tau_s-\tau_0}\right] \exp^{-\big(\frac{h_0\gamma_a}{\tau_s-\tau_0}\big)}
\end{equation}

\noindent where $h_0$ is the initial hardening modulus, $\tau_0$ the initial yield shear stress, $\tau_s$ the saturation yield shear stress, $h_s$ the saturation hardening modulus at large strains and $\gamma_a$ is the accumulated shear strain in all slip systems, which is given by

\begin{equation}
\gamma_a=\int^t_0\sum_{\alpha}|\dot{\gamma}^{\alpha}|dt.
\end{equation}

The precipitates are obstacles to the dislocation motion in each slip system and their effect should be reflected in the self-hardening modulus. Because the precipitates are disks parallel to the \{100\} faces of the Al FCC lattice, their influence is the same for all slip systems and, thus, the self-hardening modulus is equivalent for all slip systems but different for each aging condition.

\subsection{Finite element model of micropillar compression}

\begin{figure}[h!]
	\centering
	\includegraphics[height=8cm]{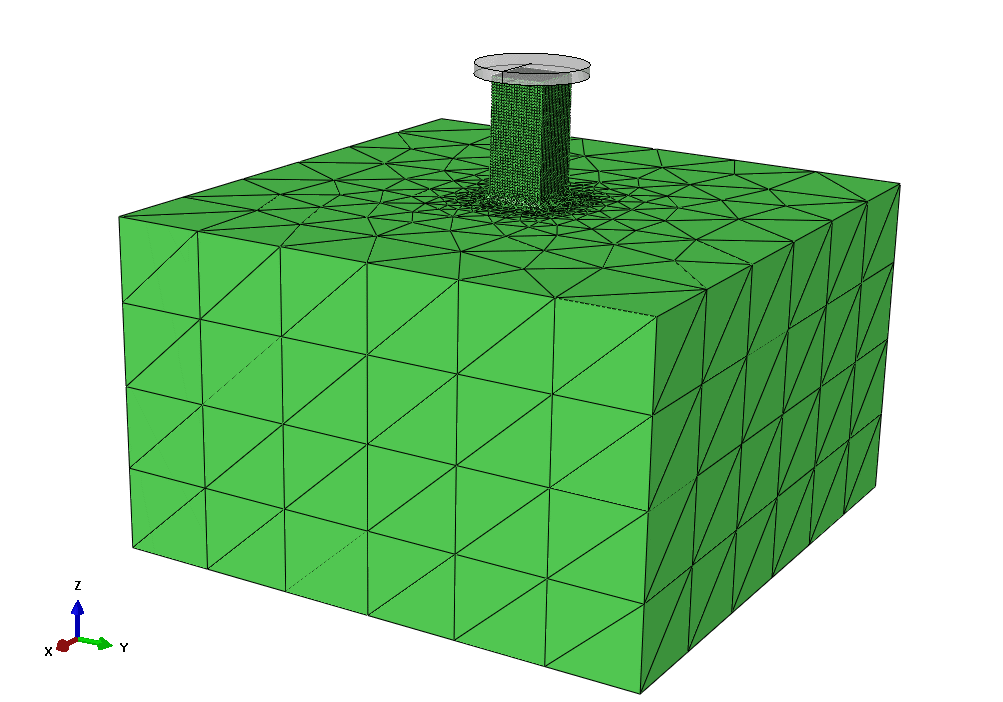}
	\caption{Finite element model of the micropillar compression test}
	\label{fig:micropillar_FEM_model}
\end{figure}

The crystal plasticity model presented above was programmed as a User Material Subroutine in Abaqus/Standard \cite{A19} and it was used to simulate the micropillar compression tests of artificially aged samples, which showed homogeneous deformation along the micropillar.  The corresponding finite element model is shown in Fig. \ref{fig:micropillar_FEM_model}. The model includes the geometry of the square micropillars with 5 $\mu $m in lateral size and the radius of curvature of the fillet (approximately 1.5 $\mu$m). The micropillar and the supporting material were modelled as a single crystal.  Eight-node quadratic elements  (C3D8) were used for the discretization of the micropillar while four-node linear elements (C3D4) were used for the supporting material. The total number of elements in the model were 31736 elements. The flat punch was modelled as a rigid body, with a lateral stiffness of 10 KN/m, which approximately corresponds to the lateral stiffness of the indenter used in the experiment \cite{Soler2012}. The contact between the flat punch and the micropillar surface was modelled according to a Coulomb friction, with a friction coefficient of 0.1 The base of the supporting material was fully constrained, while a flat punch was moved in the vertical direction at a constant speed. Simulations were carried out within the framework of the finite deformations theory with the initial unstressed state as reference.

The elastic constants of the Al single crystal elastic were obtained from DFT calculations \cite{Santos-Guemes2018} and were in good agreement with experimental data in Al single crystals at ambient temperature \cite{Kamm1964}. They can be found in Table \ref{tab:elastic-constant}. In addition, the strain rate sensitivity parameter $m$ was chosen as 0.02, following previous investigations in Al \cite{RHL19}. For each aging condition, the parameters of the Voce hardening law, eq. \eqref{eq-voceHardening}, that control the plastic response of the material were determined by comparison with the stress-plastic strain curves of micropillars oriented for single slip (Figs.  \ref{fig:OrientationAA30}a and \ref{fig:OrientationAA168}a), except $\tau_0$ that was obtained from the initial CRSS in the micropillars oriented for single slip and can be found in Table  \ref{tab:CRSS}. The other parameters, $ \tau_s $, $ h_0 $ and $ h_s $, can be found in Table \ref{tab:vocehardening}. 

\begin{table}[h!]
	\centering
	\caption{Elastic constants of the FCC Al single crystals}
	\begin{tabular}{cccc}
		\toprule
		C$ _{11} $ (GPa)& C$ _{12} $ (GPa) & C$ _{44} $ (GPa) & Ref.\\
		\midrule
		110.4  & 60.0 & 31.6 & \cite{Santos-Guemes2018}\\
		106.8  & 60.8 & 28.2 & \cite{Kamm1964} \\
		\bottomrule
	\end{tabular}%
	\label{tab:elastic-constant}%
\end{table}%

\begin{table}[htbp]
  \centering
  \caption{Parameters of the Voce hardening law (expressed in MPa) for each aging condition.}
    \begin{tabular}{lcccc}
    \toprule
          & $\tau_0$ & $\tau_s$ & $h_0$ & $h_s$  \\
          \midrule
    Artificially aged 30 h  & 90.5  & 105   & 16000    & 135 \\
    Artificially aged 168 h & 80.0    & 106.0   & 15500    & 125 \\
    \bottomrule
    \end{tabular}%
  \label{tab:vocehardening}%
\end{table}%

The 12 slip systems of the family $ \{111\} \langle 110\rangle  $ were implemented in the crystal plasticity model. 
The hardening due to the interactions among dislocations  was included in the model through the latent hardening coefficients matrix $q_{\alpha\beta}$ for each pair of slip systems. FCC crystals have 12 slip systems but only six independent coefficients are necessary to determine the 12 x 12 coefficients of the latent hardening matrix due to symmetry considerations \cite{Haouala2018}. They stand for the different types of interactions between dislocations (self interaction, coplanar and collinear) as well as locks (glissile junction, Hirth and Lomer-Cottrell lock).  They have been calculated by means of dislocation dynamics simulations for pure FCC single crystals \cite{KDH08, Bertin2013} but they are not known in the case of precipitation-hardened crystals. So, they were also obtained by comparison of the simulation results with the experimental curves for micropillars with different orientation. In particular, the self-hardening coefficient was assumed to be equal to 1 and it was used to determine the parameters in the Voce law from the micropillar compression tests in crystals oriented for single slip. The micropillar compression tests oriented for coplanar and non coplanar double slip were used to determine the coplanar and collinear coefficients, respectively. Finally, the parameters corresponding to the glissile junction and Hirth and Lomer-Cottrell locks were obtained from the tests on micropillars oriented for multiple slip. However, the Hirth and Lomer-Cottrell also influenced the mechanical behavior of  micropillars aligned near [100] and [011] orientations, respectively, and the optimum values were obtained after a few iterations. Overall, the experimental results and the finite element simulations of the mechanical response of micropillars compressed along different orientations  (Figs.  \ref{fig:OrientationAA30} and \ref{fig:OrientationAA168}) were in good agreement.

\begin{table}[h]
\centering
\caption{Parameters of the hardening interaction matrix between slip systems for each of the aging conditions.}
\begin{tabular}{lcccccc}
\toprule
 & Self & Coplanar & Collinear & Hirth & Glissile & Lomer \\ 
 \midrule
Artificially aged 30 h & 1.0 & 0.8 & 0.8 & 2.2 & 1.0 & 1.0 \\
Artificially aged 168 h & 1.0 & 0.8 & 0.8 & 3.0 & 1.0 & 1.0 \\ \bottomrule
\end{tabular}
\end{table}

\subsection{Polycrystal homogenization}

\begin{figure}[h!]
	\centering
	\includegraphics[height=8cm]{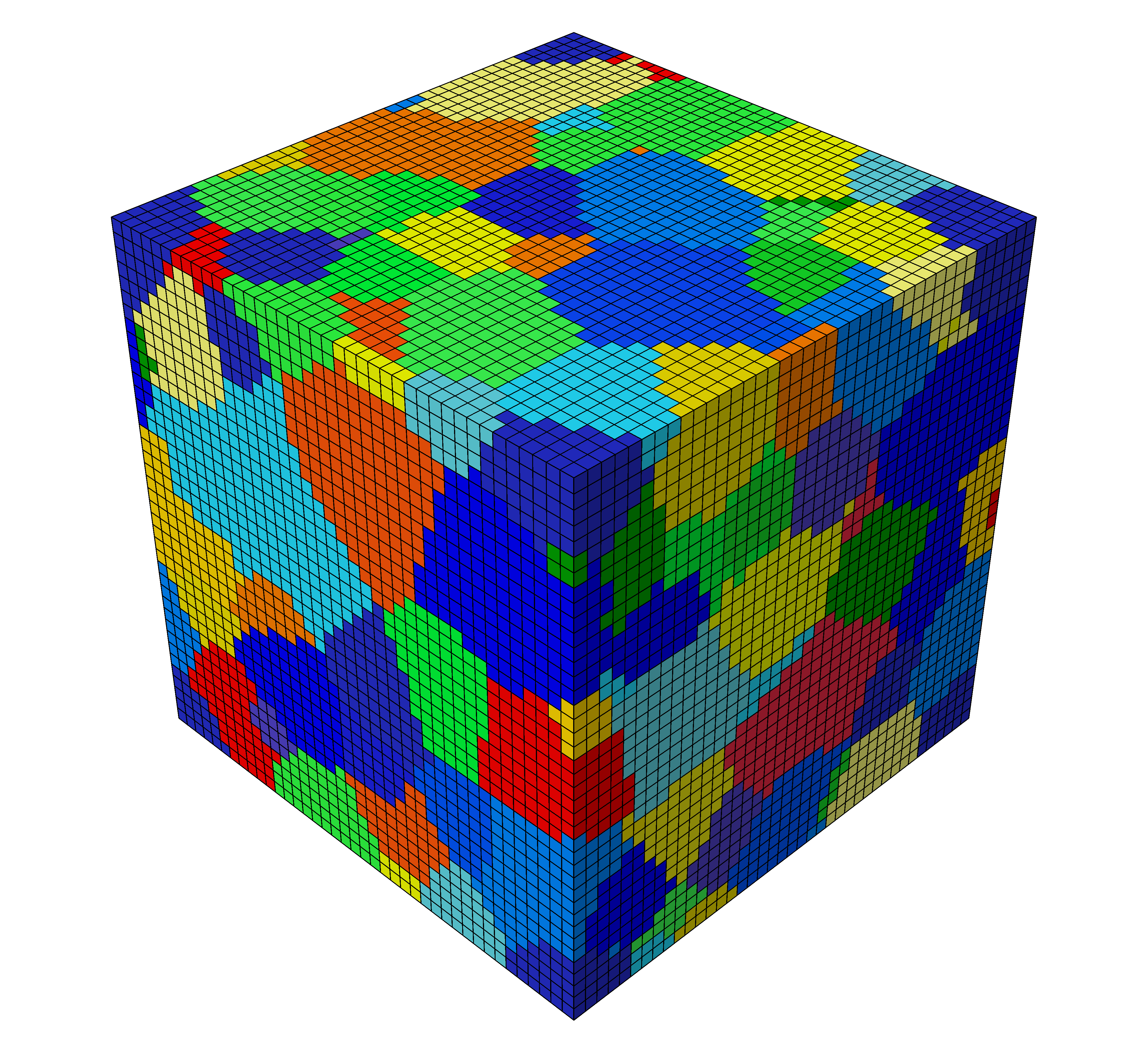}
	\caption{Representative volume element of polycrystalline Al.Cu alloy with random texture  containing 100 crystals discretized with 64000 cubic finite elements.}
	\label{fig:RVE}
\end{figure}

The mechanical properties of Al-Cu polycrystals  can be computed by means of the finite element simulation of a representative volume element  (RVE) of the microstructure using the crystal plasticity presented above \cite{Segurado2018}. The RVE was made up of 100 equiaxed grains and was generated using  Dream3D \cite{Dream3D} assuming a log-normal distribution grain size distribution with a standard deviation of 20\% of the average grain size and random texture (Fig. \ref{fig:RVE}). It should be noted that no length scale was included in the constitutive model and, thus, the results of the simulation are independent of the grain size (the strengthening due to Hall-Petch effect is not taken into account in this model). The RVE was discretized with a regular mesh of 40 x 40 x 40 quadratic cubic finite elements with full integration (C3D8 in Abaqus). The mechanical response of the polycrystals in uniaxial tension under quasi-static loading conditions was determined using periodic boundary conditions with Abaqus/Standard \cite{A19} within the framework of the finite deformations theory with the initial unstressed state as reference . More details about the computational homogenization strategy can be found in \cite{Haouala2018, RHL19}.

\begin{figure}[h!]
	\centering
	\includegraphics[height=8cm]{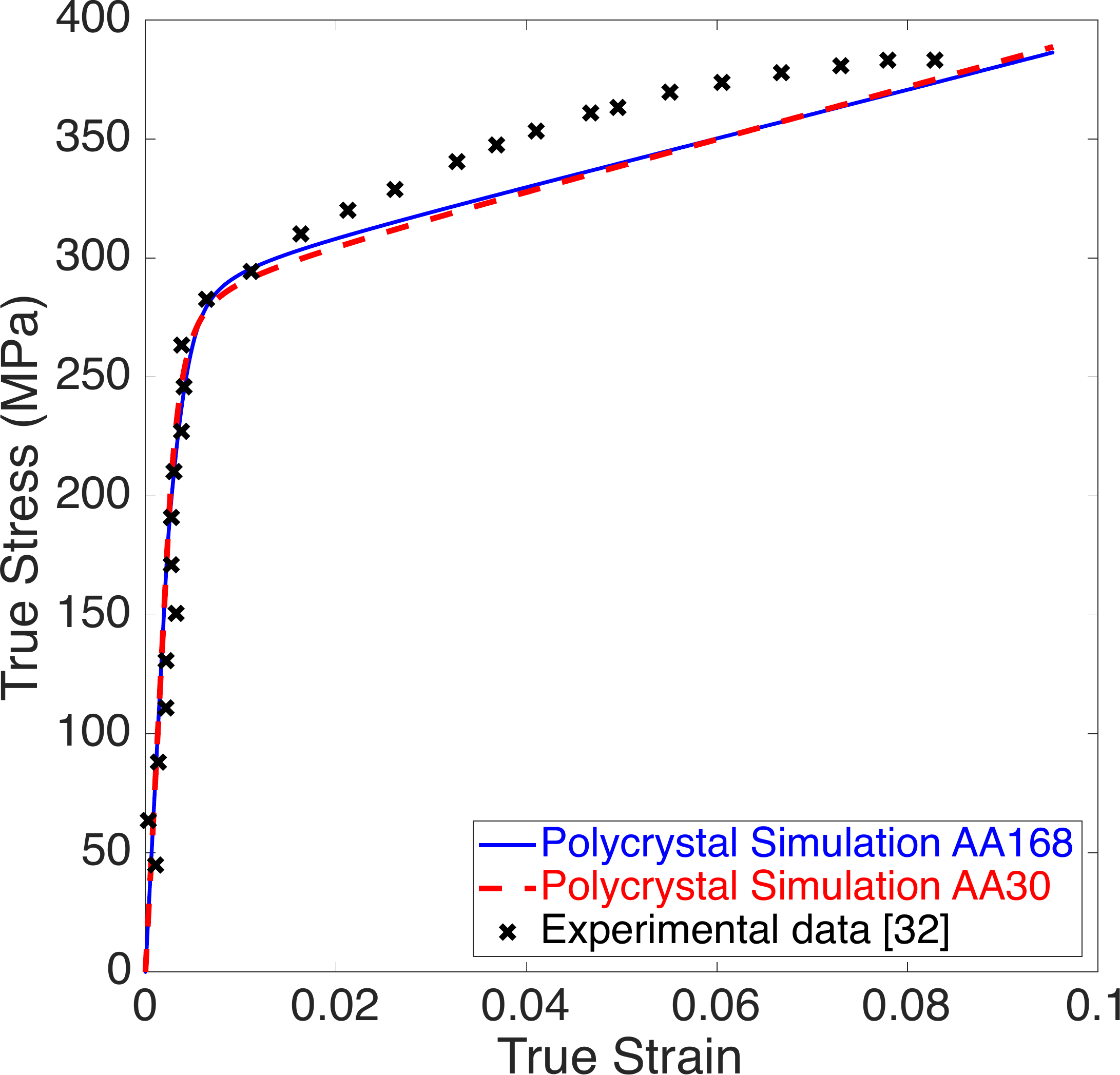}
	\caption{True tensile stress-strain curves of polycrystalline Al-4 wt. \% Cu alloys subjected to different aging treatments obtained from the numerical simulation of a representative volume of the microstructure. Experimental results of an artificially aged Al-3 wt. \%Cu alloy \cite{DaCostaTeixeira2009} is also plotted for comparison.}
	\label{fig:poly}
\end{figure}

The true stress-strain curves obtained from the polycrystal simulations are plotted in Fig. \ref{fig:poly} for the Al-4 wt. \% Cu alloys artificially aged at 180$^\circ$ during 30 and 168 h. The curves corresponding to the AA30 and AA168 alloys are practically superposed and indicate that high strength and hardening can be obtained in this alloy with two different types of precipitate structures: either a fine dispersion of $\theta''$ precipitates or a coarser one of $\theta'$ precipitates.  Experimental results of the tensile properties of Al-Cu polycrystals with similar Cu content and aging treatments are also plotted in Fig. \ref{fig:poly} for comparison. Da Costa Texeira {\it et al.} \cite{DaCostaTeixeira2009} measured the tensile deformation of an Al - 3 wt. \% Cu alloy which contained traces of Sn to favour the nucleation and growth of $\theta'$ precipitates. The alloy was solution heat treated at 525$^\circ$C, water quenched and aged at 200$^\circ$C, leading to a homogeneous distribution of $\theta'$ precipitates. The precipitate volume fraction and diameter were, respectively, 1.88 \% and 103 $\mu$m and this aging condition led to the best mechanical properties. Moreover, the dislocations overcame the precipitates by the formation of Orowan loops. No information was provided about the grain size but the influence of the Hall-Petch effect in this precipitation-hardened alloy should be minor. The experimental true stress-strain curve for this alloy is plotted in Fig. \ref{fig:poly}  and compares very well the numerical predictions for the AA30 and AA168 alloys in terms of the initial yield strength and the strain hardening. Thus, the parameters of the crystal plasticity model obtained from micropillar compression tests could be used to predict the mechanical properties of the polycrystals.

\section{Conclusions}\label{S:5}

The mechanical properties of Al - 4 wt. \% Cu alloy containing different types of precipitates (either Guinier-Preston zones, $\theta''$ or $\theta'$) were determined by means of micropillar compression tests in single crystals oriented for single, double (coplanar and non-coplanar) and multiple slip. It was found that the mechanical properties were independent of the micropillar size when the micropillar cross-section was $\ge$  5 x 5 $\mu$m$^2$. The dimensions, volume fraction and spatial distribution of the precipitates in each crystal were carefully measured by means of transmission electron microscopy, while the initial critical resolved shear stress and the hardening in different orientations were obtained from the micropillar tests, providing an unique data set to establish the relationship between the precipitate structure and the mechanical properties.

It was found that deformation was localized along the dominant deformation slip planes in the alloy reinforced with Guinier-Preston zones, as a result of precipitate shearing. Precipitate shearing was also observed in the case of   small $\theta''$ ($ <$ 30 $\mu$m in diameter) while Orowan loops were formed around larger $\theta''$ and $\theta'$ precipitates. The initial critical resolved shear stress and the hardening rate of the samples containing either $\theta''$ (aged during 30 hours at 180$^\circ$C) or  $\theta'$ (aged during 168 hours at 180$^\circ$C) precipitates were very similar and higher than those found in the naturally aged sample with Guinier-Preston zones. Thus, high strength Al-Cu alloys can be obtained through the dispersion of $\theta''$ or $\theta'$ precipitates. The high strength of the former could be traced to an Orowan mechanism and was determined by the small distance between the precipitates. The contribution of this mechanism to the strength is smaller in the case of the sample with $\theta'$ precipitates but the critical resolved shear stress was increased by the presence of the transformation strains associated to the formation of $\theta'$, as demonstrated by recent dislocation dynamics simulations \citep{SBE20}. 

Finally, the micropillar compression tests in different orientations were used to calibrate a phenomenological crystal plasticity model. This information was used to predict the mechanical properties of polycrystals by means of computational homogeneization. The simulation results were in a reasonable agreement with experimental data in the literature of Al-Cu alloys subjected to similar heat treatments and demonstrate that micropillar compression can be used to determine the mechanical properties of precipitation-hardened single crystals.

\section*{Acknowledgements}
This investigation was supported by the European Research Council under the European UnionÃs Horizon 2020 research and innovation programme (Advanced Grant VIRMETAL, grant agreement No. 669141).  BB acknowledges the support from the Spanish Ministry of Education through the Fellowship FPU15/00403. 


\begin{thebibliography}{10}
\expandafter\ifx\csname url\endcsname\relax
  \def\url#1{\texttt{#1}}\fi
\expandafter\ifx\csname urlprefix\endcsname\relax\def\urlprefix{URL }\fi
\expandafter\ifx\csname href\endcsname\relax
  \def\href#1#2{#2} \def\path#1{#1}\fi

\bibitem{Polmear2017}
I.~Polmear, D.~{St. John}, J.-F. Nie, M.~Qian, {Physical Metallurgy of
  Aluminium Alloys}, Butterworth-Heinemann, 2017.

\bibitem{Nie2014}
J.~F. Nie, Physical metallurgy of light alloys, in: Physical Metallurgy, 5th
  edition, Elsevier, 2014, pp. 2009--2156.

\bibitem{Argon2008}
A.~S. Argon, {Strengthening mechanisms in crystal plasticity}, Oxford
  University Press Oxford, 2008.

\bibitem{N97}
E.~Nembach, {Particle Strengthening of Metals and Alloys}, Jon Wiley \& Sons,
  1997.

\bibitem{Ringer2000}
S.~P. Ringer, K.~Hono, {Microstructural evolution and age hardening in
  aluminium alloys: atom probe field-ion microscopy and transmission electron
  microscopy studies}, Materials Characterization 44 (2000) 101--131.

\bibitem{Son2005}
S.~K. Son, M.~Takeda, M.~Mitome, Y.~Bando, T.~Endo, {Precipitation behavior of
  an Al-Cu alloy during isothermal aging at low temperatures}, Material Letters
  59 (2005) 629--632.

\bibitem{Guinier1938}
A.~Guinier, {Structure of age-hardened aluminium-copper alloys}, Nature
  142~(3595) (1938) 569--570.

\bibitem{Preston1938}
G.~D. Preston, {Structure of age-hardened aluminium-copper alloys}, Nature
  142~(3595) (1938) 570.

\bibitem{Silcock1953}
J.~Silcock, T.~Heal, H.~Hardy, Structural ageing characteristics of binary
  aluminium-copper alloys, Journal of the Institute of Metals 82.

\bibitem{Kaira2018}
C.~S. Kaira, C.~Kantzos, J.~J. Williams, V.~{De Andrade}, F.~{De Carlo},
  N.~Chawla, {Microstructural evolution and deformation behavior of Al-Cu
  alloys: A Transmission X-ray Microscopy (TXM) and micropillar compression
  study}, Acta Materialia 144 (2018) 419--431.

\bibitem{VS67}
D.~Vaughan, J.~M. Silcock, {The orientation and shape of $\theta$ precipitates
  formed in an Al-Cu alloy}, Physica Status Solidi 20 (1967) 725--736.

\bibitem{Boyd1971}
J.~D. Boyd, R.~B. Nicholson, {The coarsening behaviour of $\theta''$ and
  $\theta'$ precipitates in two Al-Cu alloys}, Acta Metallurgica 19 (1971)
  1379--1391.

\bibitem{Bourgeois2011}
L.~Bourgeois, C.~Dwyer, M.~Weyland, J.~F. Nie, B.~C. Muddle, {Structure and
  energetics of the coherent interface between the $\theta'$ precipitate phase
  and aluminium in Al-Cu}, Acta Materialia 59 (2011) 7043--7050.

\bibitem{Alex2018}
A.~Rodr{\'\i}guez-Veiga, B.~Bell{\'{o}}n, I.~Papadimitriou,
  G.~Esteban-Manzanares, I.~Sabirov, J.~LLorca, {A multidisciplinary approach
  to study precipitation kinetics and hardening in an Al-4Cu (wt. {\%}) alloy},
  Journal of Alloys and Compounds 757 (2018) 504--519.

\bibitem{Biswas2011}
A.~Biswas, D.~J. Siegel, C.~Wolverton, D.~N. Seidman, {Precipitates in Al-Cu
  alloys revisited: Atom-probe tomographic experiments and first-principles
  calculations of compositional evolution and interfacial segregation}, Acta
  Materialia 59 (2011) 6187--6204.

\bibitem{KPG17}
K.~Kim, A.~Roy, M.~P. Gururajan, C.~Wolverton, P.~W. Voorhees,
  {First-principles/phase-field modeling of $\theta'$ precipitation in Al-Cu
  alloys}, Acta Materialia 140 (2017) 344--354.

\bibitem{Liu2017}
H.~Liu, B.~Bell{\'{o}}n, J.~LLorca, {Multiscale modelling of the morphology and
  spatial distribution of $\theta '$ precipitates in Al-Cu alloys}, Acta
  Materialia 132 (2017) 611--626.

\bibitem{LPL19}
H.~Liu, I.~Papadimitriou, F.~Lin, J.~LLorca, {Precipitation during high
  temperature aging of Al-Cu alloys: A multiscale analysis based on first
  principles calculations}, Acta Materialia 167 (2019) 121--135.

\bibitem{SH19}
T.~Stegm\"uller, F.~Haider, Multi-scale cluster dynamics modelling of
  {Guinier-Preston} zone formation in binary al-cu alloys, Acta Materialia 177
  (2019) 240--249.

\bibitem{MKI19}
H.~Miyosi, H.~Kimizuka, A.~Ishii, S.~Ogata, Temperature-dependent nucleeation
  kinetics of {Guiner-Preston} zones in {Al-Cu} alloys. an atomistic {Kinetic
  Monte Carlo} and classical nucleation theory approach, Acta Materialia 179
  (2019) 262--272.

\bibitem{WO01}
C.~Wolverton, V.~Ozolins, {Entropically favored ordering: the metallurgy of
  Al2Cu revisited}, Physical Review Letters 86 (2001) 5518--5521.

\bibitem{Byrne1961}
J.~G. Byrne, M.~E. Fine, A.~Kelly, Precipitate hardening in an aluminum-copper
  alloy, Philosophical Magazine 6 (1961) 1119--1145.

\bibitem{EMS19}
G.~Esteban-Manzanares, a.~J.~S. E.~Mart{\'\i}nez, L.~Capolungo, J.~LLorca, An
  atomistic investigation of the interaction of dislocations with
  guinier-preston zones in {Al-Cu} alloys, Acta Materialia 162 (2019) 189--201.

\bibitem{EBM20}
G.~Esteban-Manzanares, B.~Bell\'on, a.~I.~P. E.~Mart{\'\i}nez, J.~LLorca,
  {Strengthening of Al-Cu alloys by Guinier-Preston zones: predictions from
  atomistic simulations}, Journal of the Mechanics and Physics of Solids 132
  (2020) 103675.

\bibitem{DaCostaTeixeira2008}
J.~{da Costa Teixeira}, D.~G. Cram, L.~Bourgeois, T.~J. Bastow, A.~J. Hill,
  C.~R. Hutchinson, {On the strengthening response of aluminum alloys
  containing shear-resistant plate-shaped precipitates}, Acta Materialia 56
  (2008) 6109--6122.

\bibitem{Nie2008}
J.-F. Nie, B.~C. Muddle, {Strengthening of an Al-Cu-Sn alloy by
  deformation-resistant precipitate plates}, Acta Materialia 56 (2008)
  3490--3501.

\bibitem{Sehitoglu2005}
H.~Sehitoglu, T.~Foglesong, H.~J. Maier, {Precipitate effects on the mechanical
  behavior of aluminum copper alloys: Part I. Experiments}, Metallurgical and
  Material Transactions A 36 (2005) 751--761.

\bibitem{Ma2016}
P.~P. Ma, C.~H. Liu, C.~L. Wu, L.~M. Liu, J.~H. Chen, {Mechanical properties
  enhanced by deformation-modified precipitation of $\theta'$-phase
  approximants in an Al-Cu alloy}, Materials Science and Engineering A 676
  (2016) 138--145.

\bibitem{Li2017}
S.~H. Li, W.~Z. Han, J.~Li, E.~Ma, Z.~W. Shan, {Small-volume aluminum alloys
  with native oxide shell deliver unprecedented strength and toughness}, Acta
  Materialia 126 (2017) 202--209.

\bibitem{Kaira2019}
C.~S. Kaira, T.~J. Stannard, V.~D. Andrade, F.~D. Carlo, N.~Chawla, Exploring
  novel deformation mechanisms in aluminum-copper alloys using in situ {4D}
  nanomechanical testing, Acta Materialia 176 (2019) 242 -- 249.

\bibitem{Calabrese1974}
C.~Calabrese, C.~Laird, {Cyclic stress-strain response of two-phase alloys Part
  I. Microstructures containing particles penetrable by dislocations},
  Materials Science and Engineering 13 (1974) 141--157.

\bibitem{Calabrese1974b}
C.~Calabrese, C.~Laird, {Cyclic stress-strain response of two-phase alloys Part
  II. Particles Not Penetrated by Dislocations}, Materials Science and
  Engineering 13 (1974) 159--174.

\bibitem{Merle1981a}
P.~Merle, F.~Fouquet, J.~Merlin, {Experimental and theoretical determinations
  of the yield stress of an alloy containing plate-like precipitates: $\theta
  '$ phase in an Al-4wt.{\%}Cu alloy}, Materials Science and Engineering 50
  (1981) 215--220.

\bibitem{DaCostaTeixeira2009}
J.~{da Costa Teixeira}, L.~Bourgeois, C.~W. Sinclair, C.~R. Hutchinson, {The
  effect of shear-resistant, plate-shaped precipitates on the work hardening of
  Al alloys: Towards a prediction of the strength-elongation correlation}, Acta
  Materialia 57 (2009) 6075--6089.

\bibitem{Chen2013}
Y.~Chen, M.~Weyland, C.~R. Hutchinson, {The effect of interrupted aging on the
  yield strength and uniform elongation of precipitation-hardened Al alloys},
  Acta Materialia 61 (2013) 5877--5894.

\bibitem{Price1964}
R.~J. Price, A.~Kelly, {Deformation of age-hardened aluminium alloy crystals-I
  plastic flow}, Acta Metallurgica 12 (1964) 159--169.

\bibitem{Russell1970}
K.~G. Russell, M.~F. Ashby, {Slip in Aluminum Crystals Containing Strong,
  Plate-Like Particles}, Acta Metallurgica 18 (1970) 891--901.

\bibitem{Muraishi2002}
S.~Muraishi, N.~Niwa, A.~Maekawaz, S.~Kumai, A.~Satoy, {Strengthening of Al -
  Cu single crystals by stress-oriented Guinier - Preston zones}, Philosophical
  Magazine A 82 (2002) 2755--2771.

\bibitem{Santos-Guemes2018}
R.~Santos-G{\"{u}}emes, G.~Esteban-Manzanares, I.~Papadimitriou, J.~Segurado,
  L.~Capolungo, J.~LLorca, {Discrete dislocation dynamics simulations of
  dislocation-$\theta'$ precipitate interaction in Al-Cu alloys}, Journal of
  the Mechanics and Physics of Solids 118 (2018) 228--244.

\bibitem{ImageJ12}
C.~A. Schneider, W.~S. Rasband, K.~W. Eliceiri, {NIH Image to ImageJ: 25 years
  of image analysis}, Nature Methods 9 (2012) 671--675.

\bibitem{KJB75}
M.~Kelly, A.~Jostsons, R.~G. Blake, J.~G. Napier, The determination of foil
  thickness by scanning transmission electron microscopy, Physica Status Solidi
  A 31 (1975) 771--780.

\bibitem{MTEX}
R.~Hielscher, C.~B. Silbermann, E.~Schmidl, J.~Ihlemann, {Denoising of crystal
  orientation maps}, Journal of Applied Crystallography 52 (2019) 984--996.

\bibitem{Greer2011}
J.~R. Greer, J.~T.~M. {De Hosson}, {Plasticity in small-sized metallic systems:
  Intrinsic versus extrinsic size effect}, Progress in Materials Science 56
  (2011) 654--724.

\bibitem{Soler2012}
R.~Soler, J.~M. Molina-Aldareguia, J.~Segurado, J.~Llorca, R.~I. Merino, V.~M.
  Orera, {Micropillar compression of LiF [111] single crystals: Effect of size,
  ion irradiation and misorientation}, International Journal of Plasticity 36
  (2012) 50--63.

\bibitem{Wharry2019}
J.~P. Wharry, K.~H. Yano, P.~V. Patki, Intrinsic-extrinsic size effect
  relationship for micromechanical tests, Scripta Materialia 162 (2019) 63--67.

\bibitem{Proudhon2008}
H.~Proudhon, W.~J. Poole, X.~Wang, Y.~Br{\'{e}}chet, {The role of internal
  stresses on the plastic deformation of the Al-Mg-Si-Cu alloy AA6111},
  Philosophical Magazine 88 (2008) 621--640.

\bibitem{ASMHandbookV2}
{ASM Handbook Committee}, {Properties and Selection: Nonferrous Alloys and
  Special-Purpose Materials}, ASM International, 1990.

\bibitem{Phillips1973}
V.~A. Phillips, Lattice resolution measurement of strain fields at
  {Guinier-Preston} zones of strain in {Al}-3.0 {\%} cu, Acta Metallurgica 21
  (1973) 219--228.

\bibitem{Karlik2004}
M.~Karl{\'{i}}k, A.~Bigot, B.~Jouffrey, P.~Auger, S.~Belliot, {HREM, FIM and
  tomographic atom probe investigation of Guinier-Preston zones in an Al-1.54
  at{\%} Cu alloy}, Ultramicroscopy 98 (2004) 219--230.

\bibitem{Greenwood1997}
N.~Greenwood, A.~Earnshaw, {Copper, Silver and Gold}, Butterworth-Heinemann,
  1997, pp. 1173 -- 1200.

\bibitem{Cruzado2015}
A.~Cruzado, B.~Gan, M.~Jim{\'{e}}nez, D.~Barba, K.~Ostolaza, A.~Linaza, J.~M.
  Molina-Aldareguia, J.~Llorca, J.~Segurado, {Multiscale modeling of the
  mechanical behavior of IN718 superalloy based on micropillar compression and
  computational homogenization}, Acta Materialia 98 (2015) 242--253.

\bibitem{T34}
G.~I. Taylor, The mechanism of plastic deformation of crystals, Proceedings of
  the Royal Society {A}165 (1934) 362--387.

\bibitem{RHL19}
R.~A. Rubio, S.~Haouala, J.~LLorca, Grain boundary strengthening of {FCC}
  polycrystals, Journal of Materials Research 34 (2019) 2263 -- 2274.

\bibitem{KM03}
U.~F. Kocks, H.~Mecking, Physics and phenomenology of strain hardening: the
  {FCC} case, Progress in Materials Science 48 (2003) 171 -- 273.

\bibitem{ZS08}
Z.~Zhu, M.~J. Starink, Age hardening and softening in cold-rolled {Al-Mg-Mn
  alloys with up to 0.4wt\% Cu}, Materials Science and Engineering A 489 (2008)
  138 -- 149.

\bibitem{SBE20}
R.~Santos-G{\"{u}}emes, B.~Bell\'on, G.~Esteban-Manzanares, J.~Segurado,
  L.~Capolungo, J.~LLorca, Multiscale modelling of precipitation hardening in
  {Al-Cu} alloys: dislocation dynamics simulations and experimental validation,
  Acta Materialia 188 (2020) 475 -- 485.

\bibitem{DW83}
D.~Dahmen, K.~H. Westmacott, The mechanisms of $\theta'$ precipitation on
  climbing dislocations on {Al-Cu}, Scripta Metallurgica 17 (1983) 1241--1246.

\bibitem{JZJ05}
Z.~Jiang, Q.~Zhang, H.~Jiang, Z.~Chen, X.~Wu, Spatial characteristics of the
  portevin-le chatelier deformation bands in {Al-4at\%Cu }polycrystals,
  Materials Science and Engineering A 403 (2005) 154 -- 164.

\bibitem{JZC07}
H.~Jiang, Q.~Zhang, X.~Chen, Z.~Chen, Z.~Jiang, X.~Wu, J.~Fan, Three types of
  {Portevin-Le Chatelier effects: Experiment and modelling}, Acta Materialia 55
  (2007) 2219 -- 2228.

\bibitem{Segurado2018}
J.~Segurado, R.~A. Lebensohn, J.~Llorca, {Computational Homogenization of
  Polycrystals}, Advances in Applied Mechanics 51 (2018) 1--114.

\bibitem{A19}
Abaqus, Analysis {U}ser's manual, Dassault Syst\`emes.

\bibitem{Kamm1964}
G.~N. Kamm, G.~A. Alers, {Low-temperature elastic moduli of aluminum}, Journal
  of Applied Physics 35 (1964) 327--330.

\bibitem{Haouala2018}
S.~Haouala, J.~Segurado, J.~LLorca, {An analysis of the influence of grain size
  on the strength of FCC polycrystals by means of computational
  homogenization}, Acta Materialia 148 (2018) 72--85.

\bibitem{KDH08}
L.~Kubin, B.~Devincre, T.~Hoc, Modeling dislocation storage rates and mean free
  paths in face-centered cubic crystals., Acta Materialia 56 (2008) 6040--6049.

\bibitem{Bertin2013}
N.~Bertin, L.~Capolungo, I.~J. Beyerlein, {Hybrid dislocation dynamics based
  strain hardening constitutive model}, International Journal of Plasticity 49
  (2013) 119--144.

\bibitem{Dream3D}
M.~A. Groeber, M.~A. Jackson, {DREAM.3D: A Digital Representation Environment
  for the Analysis of Microstructure in 3D}, Integrating Materials and
  Manufacturing Innovation 3 (2014) 56--72.

\end{thebibliography}

\end{document}